\documentclass[aps,prc, dvips, twocolumn,groupedaddress,showkeys,showpacs,floatfix,superscriptaddress]{revtex4-1}
\usepackage{subfigure}
\usepackage{color,graphicx}
\usepackage{amsmath,amssymb}
\usepackage{enumerate}
\usepackage{bm}
\usepackage[%
bookmarks=true,%
bookmarksnumbered=true,%
bookmarkstype=toc,%
colorlinks=true,%
linkcolor=blue,%
citecolor=blue,%
]{hyperref}
\newcommand{\<}{\langle}
\renewcommand{\>}{\rangle}

\begin{document}

\title{Introduction of the one-body correlation operator in the unitary-model-operator approach}
\author{Takayuki Miyagi}
\affiliation{Center for Nuclear Study, the University of Tokyo, Hongo, Tokyo 113-0033, Japan}
\author{Takashi Abe}
\affiliation{Department of Physics, the University of Tokyo, Hongo, Tokyo 113-0033, Japan}
\author{Ryoji Okamoto}
\affiliation{Senior Academy, Kyushu Institute of Technology, Tobata, Kitakyushu 804-0015, Japan}
\author{Takaharu Otsuka}
\affiliation{RIKEN Nishina Center, 2-1 Hirosawa, Wako, Saitama 351-0198, Japan}
\affiliation{Department of Physics, the University of Tokyo, Hongo, Tokyo 113-0033, Japan}
\affiliation{National Superconducting Cyclotron Laboratory, Michigan State University, East Lansing, Michigan 48824, USA}
\affiliation{Instituut voor Kern- en Stralingsfysica, KU Leuven, B-3001 Leuven, Belgium}

\date{\today}

\begin{abstract}
In the earlier unitary-model-operator approach (UMOA),
one-body correlations have been taken into account
 approximately by the diagonalization of unitary-transformed Hamiltonians in the $0p0h$ and $1p1h$ space.
With this prescription, the dependence of the harmonic-oscillator energy ($\hbar\omega$)
on calculated observables is not negligible even at larger model spaces.
In the present work, we explicitly introduce the one-body correlation operator
so that it optimizes the single-particle basis states and then reduces the $\hbar\omega$-dependence.
For an actual demonstration, we calculate the energy and radius for the $^{4}$He ground state
with the softened nucleon-nucleon ($NN$) interactions from Argonne v18 (AV18)
and chiral effective field theory ($\chi$EFT) up to the next-to-next-to-next leading order (N$^{3}$LO).
As a result, we obtain practically $\hbar\omega$-free results at sufficiently large model spaces.
The present results are reasonably close to those by the other {\em ab initio} calculations with the same $NN$ interactions.
This methodological development enables us more systematic analysis of calculation results in the UMOA.
We also discuss qualitatively the origin of the $\hbar\omega$-dependence on calculated observables
in a somewhat simplified way.
\end{abstract}

\pacs{21.10.Dr,21.60.De}
\maketitle

\section{\label{sec:intro}Introduction}
Over the past decade, nuclear structure for medium-mass nuclei
 have been extensively investigated by {\it ab inito} many-body approaches such as
 the coupled-cluster method (CCM)~\cite{Wloch:2005,Hagen:2007,Hagen:2008,Hagen:2010,
Hagen:2012-1,Hagen:2012-2,Binder:2013,Hagen:2014,Binder:2015},
in-medium similarity renormalization group (IM-SRG)
 approach~\cite{Hergert:2013-1,Hergert:2013-2,Hergert:2014},
and self-consistent Green's function method~\cite{Barbieri:2009,Cipollone:2013,Soma:2014}.
For instance, the reproduction of ground-state energies for oxygen isotopes is one of the most
 successful examples with such methods~\cite{Hagen:2012-1, Hergert:2013-2, Cipollone:2013}.
This finding supports the importance of the contributions
from three-body forces for exotic nuclear structure.
Also, the CCM casts light on the size of atomic nucleus from the study of
neutron and weak-charge distributions of the $^{48}$Ca nucleus~\cite{Hagen:2015gr}.
In addition, the IM-SRG stimulates discussion on nuclear shape deformations
with such {\it ab initio} approaches~\cite{Stroberg:2016jf},
which follows the extension of the method to solve the valence-space problem
for open-shell nuclei~\cite{Tsukiyama:2012dw,Bogner:2014ib}.

Similarly to these methods, the unitary-model-operator approach (UMOA)~\cite{Suzuki:1994, Fujii:2004, Fujii:2009}
 is also applicable to the investigation for the medium-mass nuclei.
The UMOA was firstly introduced by Provid\^encia and Shakin to study the effects of
  short-range correlations on nuclear wave functions~\cite{Providencia:1964, Shakin:1967}.
In those studies, the correlation operator was empirically determined.
After these exploratory studies, the determination of correlation operators has been established
 based on a Hermitian effective-interaction theory~\cite{Suzuki:1986, Suzuki:1987}.
Applying this on many-body systems,
 the Okubo-Lee-Suzuki transformation is done for many-body Hamiltonians to decouple the $n$-particle $n$-hole ($npnh$) excitations.
In other words, the UMOA is natural extension of the Hartree-Fock (HF) method
where only the $1p1h$ decoupling is considered~\cite{Suzuki:1988}.
For actual applications of the UMOA, only the $2p2h$ excitations have been decoupled with the $0p0h$ state.
As mentioned in our previous publications~\cite{Miyagi:2014, Miyagi:2015},
the results strongly depend on the harmonic-oscillator-energy $\hbar\omega$,
 even if these are converged with respect to the size of model space.
It can be an issue to obtain reliable converged results
which should be free from the underlying parameters (the model-space sizes and $\hbar\omega$ values)
and also to compare with the other {\it ab initio} results.

According to the recent CCM study~\cite{Kohno:2012}, the $\hbar\omega$-dependence can be
drastically reduced by considering the one-body correlations.
In the present work, we follow this study and introduce the one-body correlation operator in the UMOA
in addition to the two-body correlation operator included already in the former calculations.
We expect that the one-body correlation operator optimizes single-particle states and
 controls the $\hbar\omega$-dependence on calculated observables.
The main purpose of this work is to demonstrate how the introduction of the one-body correlation operator
in the UMOA works well and to discuss the $\hbar\omega$-dependence of the ground-state
 energy and point-nucleon radius for the $^4$He nucleus taken as a test case.

The outline of this paper is as follows.
In Sec. \ref{sec:umoa}, we represent the theoretical framework of the UMOA
focusing on the difference between the previous and current calculation procedures.
Then, we show the results of ground-state energies and point-nucleon radii for $^{4}$He
to check how the $\hbar\omega$-dependence on these observables
can be removed by the implementation of the one-body correlation operator in Sec. \ref{sec:res}.
We also compare the UMOA results with the other {\it ab initio} ones with the same interactions.
Finally, we summarize the present work in Sec. \ref{sec:sum}.

\section{\label{sec:umoa}Unitary-model-operator approach}
Before discussing how the $1p1h$ decoupling works well in the UMOA,
we present theoretical structure in this section.
In Sec.~\ref{sec:sim}, general transformed Hamiltonians in the UMOA are introduced
to help the understanding of the difference between the formulations employed
in the earlier and this studies.
Then, it is referred how to truncate the model space in the UMOA in Sec.~\ref{sec:model}.
Sec.~\ref{sec:cal} describes the actual calculation procedure
focusing on the difference between the formulations employed in the former and this studies.
 Finally, in Sec.~\ref{sec:ob}, we mention how to calculate the observables in the UMOA.

 \subsection{Similarity transformation\label{sec:sim}}
We begin with the many-body Schr\"odinger equation,
 \begin{equation}
   \label{sch_eq_ori}
   H|\Psi\> = E_{\rm g.s.}|\Psi\>
 \end{equation}
 with the ground-state energy eigenvalue $E_{\rm g.s.}$ and eigenvector $|\Psi\>$.
Here, the operator $H$ is a general intrinsic Hamiltonian,
\begin{equation}
 H= \sum_{i}^{A}\frac{\mathbf{p}^{2}_{i}}{2m} - T_{\text{c.m.}} + \sum_{i<j}^{A}
  V_{ij} + \sum_{i<j<k}^{A} V_{ijk} + \cdots
\label{H_ori}
\end{equation}
with the mass number $A$, the momentum of the $i$th nucleon $\mathbf{p}_{i}$,
 the nucleon mass $m$,
 the $NN$ interaction $V_{ij}$, and the three-nucleon interaction $V_{ijk}$.
The center-of-mass kinetic energy $T_{\text{c.m.}}$
can be described with the one- and two-body terms as $T_{\text{c.m.}} =
 \sum_{i}^{A}\frac{\mathbf{p}^{2}_{i}}{2Am} +
 \sum_{i<j}^{A}\frac{\mathbf{p}_{i} \cdot \mathbf{p}_{j}}{Am}$.
Then, the Hamiltonian can be rewritten as
\begin{equation}
 \label{H_ori2}
  H=\sum_{i}^{A}t_{i}+\sum_{i<j}^{A}v_{ij} + \sum_{i<j<k}^{A}v_{ijk} + \cdots
\end{equation}
with the one-, two-, and three-body terms,
$t_{i}=\frac{A-1}{A}\frac{\mathbf{p}_{i}^{2}}{2m}$,
$v_{ij} = V_{ij} - \mathbf{p}_{i} \cdot \mathbf{p}_{j} / Am$,
and $v_{ijk} = V_{ijk}$, respectively.

To decouple the $0p0h$ state with $npnh$ states,
 the similarity transformation of the original Hamiltonian can be done as
\begin{equation}
\label{sim_tra}
 \widetilde{H}=U^{\dag}(H+W)U - U^{\dag}WU
\end{equation}
with the unitary operator $U$ and the auxiliary potential $W$.
Here, the auxiliary potential is introduced as
\begin{equation}
 \label{def_W}
  W=\sum_{i}^{A}w_{i} + \sum_{i<j}^{A}w_{ij}+\sum_{i<j<k}^{A} w_{ijk}+ \cdots
\end{equation}
so as to take into account the in-medium effects.
The operator $w_{i_{1}\cdots i_{n}}$ is the $n$-body auxiliary potential.
So far, the auxiliary potential $w_{i_{1} \cdots i_{n}}$ can be taken arbitrarily,
but is determined self-consistently as discussed in the end of Sec. \ref{sec:sim}.
 With the transformation (\ref{sim_tra}), the original Schr\"odinger
 equation, Eq. (\ref{sch_eq_ori}), is also transformed to
 \begin{equation}
   \widetilde{H} |\Phi\> = E |\Phi\>
 \end{equation}
with the reference state $|\Phi\>$ multiplied by the unitary operator $U$ as
\begin{equation}
  |\Phi\> = U^{\dag} |\Psi\>.
\end{equation}
Since the reference state is arbitrary in principle,
we take $|\Phi\>$ as a single Slater determinant
 such as the particle-hole vacuum.
In the UMOA, the unitary-transformation operator $U$ is defined as the product
of exponential operators up to the $A$-body terms \cite{Suzuki:1988},
\begin{equation}
\label{uni_op}
 U=e^{S^{(1)}}e^{S^{(2)}}\cdots e^{S^{(A)}}.
\end{equation}
The exponents $S^{(1)}$, $S^{(2)}$, $\dots$, $S^{(n)}$ are
the one-, two-, $\dots$, and $n$-body correlation operators.
They are defined as
\begin{align}
 S^{(1)} &= \sum_{i}^{A}s_{i}, \\
 S^{(2)} &= \sum_{i<j}^{A} s_{ij}, \\
 S^{(n)} &= \sum_{i_{1}<\cdots < i_{n}}^{A} s_{i_{1}\cdots i_{n}},
\end{align}
respectively.
Here, $s_{i_{1}\cdots i_{n}}$ is the correlation operator
 acting on $n$ particles labeled by $i_{1}, \cdots, i_{n}$.
The correlation operators $S^{(n)}$ are anti-Hermitian and satisfy
\begin{equation}
\label{S_con}
 S^{(n)\dag} = -S^{(n)},
\end{equation}
so that the transformation operator $U$ is unitary.

\begin{figure*}[t]
\subfigure[]{
  \includegraphics[keepaspectratio, width=12cm, clip]{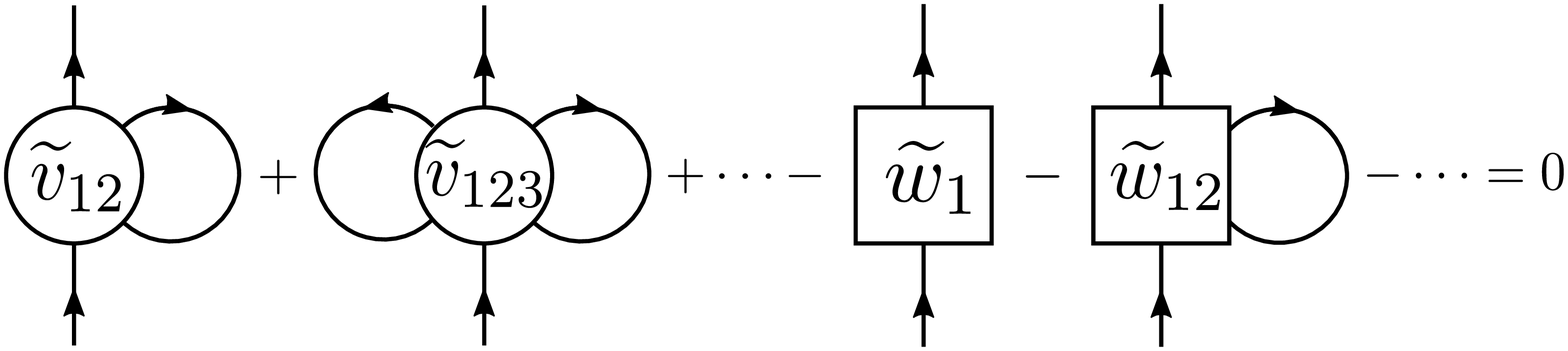}
\label{one_bubble}}
\subfigure[]{
\includegraphics[keepaspectratio, width=12cm, clip]{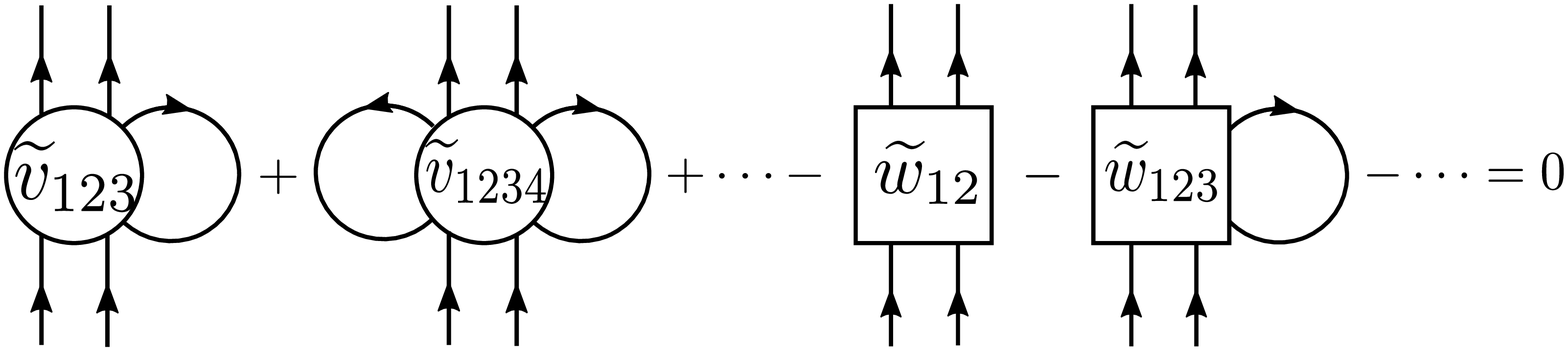}
\label{two_bubble}}
\caption{\label{bubble:fig}Cancellations of bubble-diagram contributions for the one-body (a) and two-body (b) parts.}
\end{figure*}

Generally, the transformed Hamiltonian $\widetilde{H}$ is expanded by the
 Baker-Campbell-Hausdorff (BCH) formula as found in the CCM \cite{Hagen:2014}.
It is because the BCH expansion terminates with the finite order and
 is actually one of the advantages in the CCM.
In contrast, the BCH expansion does not terminate with
 the finite order in the UMOA, as the correlation operators
 do not commute with each other.
Therefore, the UMOA employs another type of expansion known as the
 cluster expansion \cite{Providencia:1964}.
Following the cluster expansion, we decompose $\widetilde{H}$ into
\begin{equation}
  \label{cl_ex}
 \widetilde{H}=\widetilde{H}^{(1)}+\widetilde{H}^{(2)} +
  \widetilde{H}^{(3)} + \cdots,
\end{equation}
according to the number of interacting particles.
Note that the three- and higher-body terms can be induced by the
transformation, even if the initial Hamiltonian includes up to
the two-body interaction.
For clarification, we show the explicit expressions of the
one-, two-, and $n$-body cluster terms:
\begin{align}
\label{1b_h}
 \widetilde{H}^{(1)} &= \sum_{i}^{A}\widetilde{h}_{i},\\
\label{2b_h}
\widetilde{H}^{(2)} &=  \sum_{i<j}^{A}\widetilde{v}_{ij} - \sum_{i}^{A}\widetilde{w}_{i},\\
\label{nb_h}
\widetilde{H}^{(n)} &=\sum_{i_{1}<\cdots <i_{n}}^{A}
 \widetilde{v}_{i_{1}\cdots i_{n}}- \sum_{i_{1}<\cdots <
 i_{n-1}}^{A}\widetilde{w}_{i_{1}\cdots i_{n-1}} \\
 & \hspace{10em} \text{for } 3 \le n \le A \notag
\end{align}
with the terms introduced as
\begin{widetext}
\begin{align}
\label{1b_int}
 \widetilde{h}_{1}&= e^{-s_{1}}h_{1}e^{s_{1}}=
 e^{-s_{1}}( t_{1}+w_{1})e^{s_{1}}, \\
\label{2b_int}
\widetilde{v}_{12}&= e^{-s_{12}}e^{-(s_{1}+s_{2})} (h_{1}+h_{2}+v_{12}+w_{12})
 e^{s_{1}+s_{2}}e^{s_{12}}- (\widetilde{h}_{1}+\widetilde{h}_{2}), \\
\label{nb_int}
 \widetilde{v}_{i_{1}\cdots i_{n}} &= e^{-s_{1\cdots n}} \cdots
 e^{-(\sum_{i<j}s_{ij})}e^{-(\sum_{i}s_{i})}
 \left(\sum_{i}^{n}h_{i}+\sum_{k=2}^{n}\sum_{i<\cdots
 <i_{k}}^{n}v_{i_{1}\cdots i_{k}}+
\sum_{k =2}^{n}\sum_{i_{1}<\cdots< i_{k}}^{n}w_{i_{1}\cdots i_{k}}\right)
 \notag \\
&\hspace{5em}\times
   e^{\sum_{i}s_{i}} e^{\sum_{i<j}s_{ij}} \cdots e^{s_{1 \cdots n}}
-\left(\sum_{i}^{n}\widetilde{h}_{i}+\sum_{k=2}^{n-1}\sum_{i_{1}<\cdots <i_{k} }^{n}
 \widetilde{v}_{i_{1}\cdots i_{k}}\right), \quad \text{for } 3 \le n \le A.
\end{align}
\end{widetext}
The transformed auxiliary potentials
 $\widetilde{w}_{1}$, $\widetilde{w}_{12}$,
 $\dots$ in Eqs. (\ref{1b_h}) - (\ref{nb_h}) are, in principle, arbitrary,
 but the determinations of them are crucial in the actual calculation.
In order to determine these transformed auxiliary potentials, we recall
 the one-body potential appeared in the HF method.
In the HF calculations, the one-body potential cancels with the
 bubble-diagram contributions of the two-body interaction.
This procedure is applied directly to the UMOA.
Since the transformed Hamiltonian contains many-body transformed
 interactions and auxiliary potentials, the bubble-diagram contributions come from
 $\widetilde{v}_{12}$, $\widetilde{v}_{123}$, $\cdots$ and
 $\widetilde{w}_{1}$, $\widetilde{w}_{12}$, $\cdots$.
 The conditions of the cancellation can be represented diagrammatically in Fig. \ref{bubble:fig}.
The analytical expressions corresponding to Fig. \ref{bubble:fig} are
\begin{widetext}
\begin{equation}
 \sum_{\lambda\le \rho_{F}}
  \<\alpha\lambda|\widetilde{v}_{12}|\beta\lambda\> + \frac{1}{2!}
  \sum_{\lambda\mu \le \rho_{F}}
  \<\alpha\lambda\mu|\widetilde{v}_{123}|\beta\lambda\mu\> + \cdots
  - \<\alpha|\widetilde{w}_{1}|\beta\> -
\sum_{\lambda\le
 \rho_{F}}\<\alpha\lambda|\widetilde{w}_{12}|\beta\lambda\> - \cdots = 0
\end{equation}
for the one-body term and
\begin{equation}
\sum_{\lambda\le \rho_{F}}
  \<\alpha\beta\lambda|\widetilde{v}_{123}|\gamma\delta\lambda\> + \frac{1}{2!}
  \sum_{\lambda\mu \le \rho_{F}}
  \<\alpha\beta\lambda\mu|\widetilde{v}_{1234}|\gamma\delta\lambda\mu\> + \cdots
- \<\alpha\beta|\widetilde{w}_{12}|\gamma\delta\> -
\sum_{\lambda\le
 \rho_{F}}\<\alpha\beta\lambda|\widetilde{w}_{123}|\gamma\delta\lambda\> - \cdots = 0
\end{equation}
for the two-body term.
Here, $\rho_{F}$ denotes the Fermi level and $|\alpha_{1}\cdots
 \alpha_{n}\>$ is antisymmetrized and normalized $n$-body state.
The conditions of the cancellation for three- and higher-body terms are
 given in the same way.
Thus, the matrix elements of the transformed auxiliary potentials are
 \cite{Suzuki:1988,Suzuki:1992}
\begin{equation}
\label{aux_pot1}
 \<\alpha|\widetilde{w}_{1}|\beta\> = \sum_{\lambda_{1}\le \rho_{F}} \<
  \alpha \lambda_{1}|\widetilde{v}_{12}|\beta\lambda_{1}\> -
  \frac{1}{2!} \sum_{\lambda_{1}\lambda_{2}\le \rho_{F}}
\<\alpha
\lambda_{1}\lambda_{2}|\widetilde{v}_{123}|\beta\lambda_{1}\lambda_{2}\>
+ \cdots
\end{equation}
for the one-body potential, and
\begin{equation}
 \label{aux_pot2}
\<\alpha\beta|\widetilde{w}_{12}|\gamma\delta\> =
\sum_{\lambda_{1}\le \rho_{F}} \<\alpha\beta\lambda_{1}
|\widetilde{v}_{123}|\gamma\delta \lambda_{1}\>
- \frac{1}{2!} \sum_{\lambda_{1}\lambda_{2}\le\rho_{F}}
\<\alpha\beta\lambda_{1}\lambda_{2}|
\widetilde{v}_{1234}|\gamma\delta\lambda_{1}\lambda_{2}\> + \cdots
\end{equation}
for the two-body potential.
\end{widetext}
Furthermore, the auxiliary potentials $w_{1}$, $w_{12}$, $\dots$ in
 Eqs. (\ref{1b_int}) - (\ref{nb_int}) are related to the transformed
 auxiliary potentials $\widetilde{w}_{1}$, $\widetilde{w}_{12}$, $\dots$
  through the relevant inverse transformations
 \cite{Suzuki:1988,Suzuki:1992},
\begin{align}
 w_{1}&=e^{s_{1}}\widetilde{w}_{1}e^{-s_{1}}, \\
 w_{12} &=e^{s_{1}+s_{2}}e^{s_{12}}(\widetilde{w}_{1}+\widetilde{w}_{2}
 + \widetilde{w}_{12})e^{-s_{12}}e^{-(s_{1}+s_{2})} \notag \\
 &\hspace{7em} - (w_{1}+w_{2}).
\end{align}
As found in Refs \cite{Suzuki:1988,Suzuki:1992,Kohno:2012}, this choice of the transformed
 auxiliary potentials gives the normal ordered $\widetilde{H}$ with respect to $|\Phi\>$.

The essential point in the UMOA is to determine the correlation operators.
These are determined so that the transformed Hamiltonian does not
 induce the particle-hole excitations.
 There are a number of studies about the correlation operators (see, for example, Refs. \cite{Shavitt:1980, Westhaus:1981, Suzuki:1982}).
 For brevity, the determination of correlation operators is given in Appendix \ref{sec:dec_eq}.
Once the correlation operators are determined,
one can build up the transformed Hamiltonian with Eqs. (\ref{cl_ex}) - (\ref{nb_int}).

\subsection{Model space \label{sec:model}}
\begin{figure}[t]
\subfigure[]{
\includegraphics[clip,width=3.0in, clip]{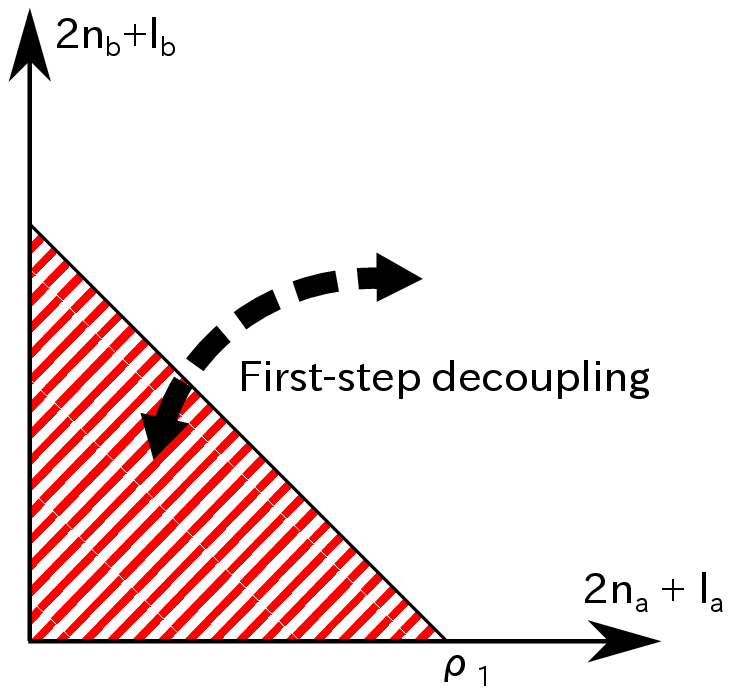}
\label{tri_1st}}
\subfigure[]{
\includegraphics[clip,width=3.0in, clip]{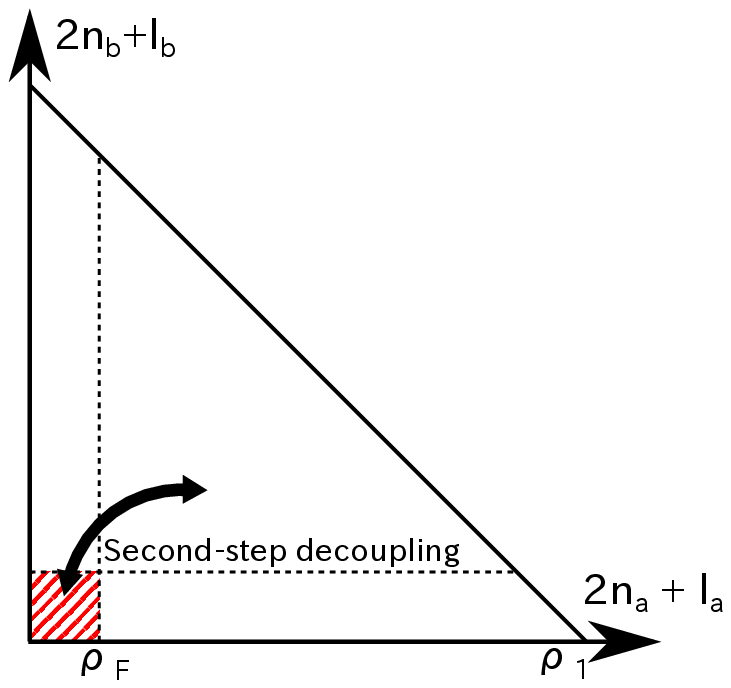}
\label{tri_2nd}}
\caption{\label{tri}(color online) Illustrations of the model space and its decoupling employed in the previous work.
 The first-step decoupling (a) denotes the decoupling between the outside and inside (shaded area) of
  our model space.
The second-step decoupling (b) means the decoupling of $2p2h$ excitations with $0p0h$ state (shaded area).}
\end{figure}

\begin{figure}[t]
\includegraphics[clip,width=3.5in]{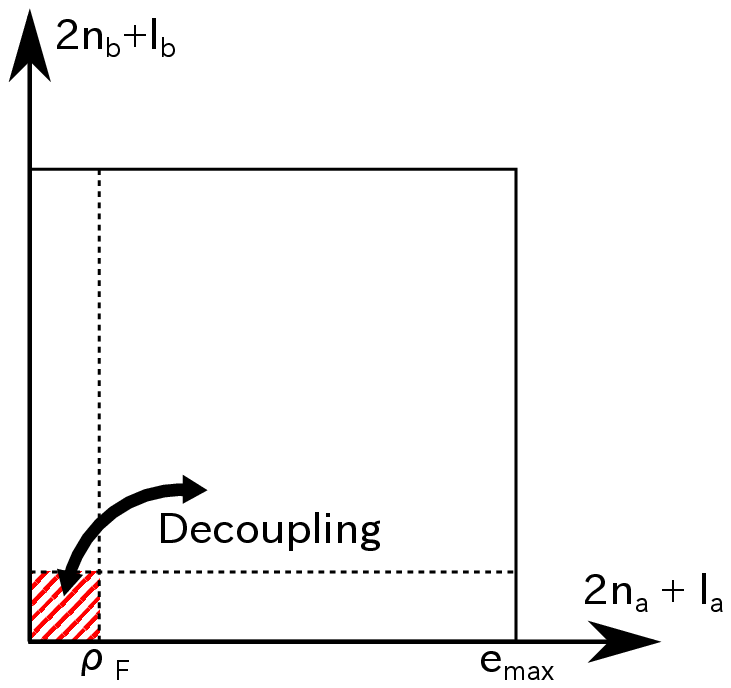}
\caption{\label{squ}(color online) Illustrations of the model space and its decoupling employed in the present work.
 The shaded area indicates our reference state. The arrow means the decoupling of $1p1h$
  and $2p2h$ excitations with the reference state.}
\end{figure}
In this subsection, we present the choice of the model space employed in earlier and current UMOA calculations.
To obtain converged results in relatively smaller mode spaces, the two-step
 decoupling method has been applied in the earlier UMOA~\cite{Fujii:2004,Fujii:2009}.
 Figs. \ref{tri_1st} and \ref{tri_2nd} schematically show how to decouple the model space in the earlier calculations.
In the first-step decoupling, the initial Hamiltonian is transformed
so that the model space (the shaded area in Fig.~\ref{tri_1st}) and its complement are decoupled.
The aim of this step is to make bare $NN$ interactions can be handled without any explicit softening of input interactions.
Since this decoupling is done in the huge space to take into
 account short-range correlations of input bare $NN$ interactions, the
 decoupling equation is solved with the relative and center-of-mass coordinates.
Thus, our model space has to be truncated to be a triangle shape in the two-body space by
 $\rho_{1}=\text{max}(2n_{a}+l_{a}+2n_{b}+l_{b})$.
Here, $n_{a}$ and $l_{a}$ are the nodal and azimuthal quantum number of the
 HO basis state $a$, respectively.
By employing this effective interaction through the first-step decoupling,
 we further decouple the $2p2h$ excitations with the reference state, which is
 illustrated by the solid arrow in Fig.~\ref{tri_2nd}.
After the second-step decoupling, we construct the transformed Hamiltonian
 and obtain the observables using the
 transformed operators as discussed in Sec.~\ref{sec:ob}.

In the first-step decoupling, the angle-average approximation is
 used for the Pauli exclusion operator.
The investigation of this approximation, at least for finite nuclei, may not be sufficient
 and can cause uncontrollable uncertainties.
 As discussed in Refs. \cite{Schiller:1999, Suzuki:2000, Baardsen:2013},
it was reported that there is the
  non-negligible difference between the results with and without the angle-average approximation
  in the nuclear matter calculations.
Since we would like to examine the validity of the UMOA without any uncontrollable approximations,
 we do not employ the first-step decoupling in the present work.
Alternatively, we soften input bare interactions via low-momentum or similarity renormalization group techniques.
Therefore, we consider only the process for the decoupling of the $1p1h$ and $2p2h$ excitations on top of the reference state.
Then, the choice of the model space is no longer restricted to the triangle shape.
Here, we employ the simplest square model space defined by
$e_{{\rm max}} = {\rm max}(2n_{a} + l_{a}) = {\rm max}(2n_{b}+l_{b})$, as shown in Fig.~\ref{squ}.
The decoupling of the $1p1h$ and $2p2h$ excitations is indicated by the solid arrow in Fig.~\ref{squ}.

\subsection{Numerical implementation \label{sec:cal}}

\begin{figure}[t]
\includegraphics[clip,width=3.5in]{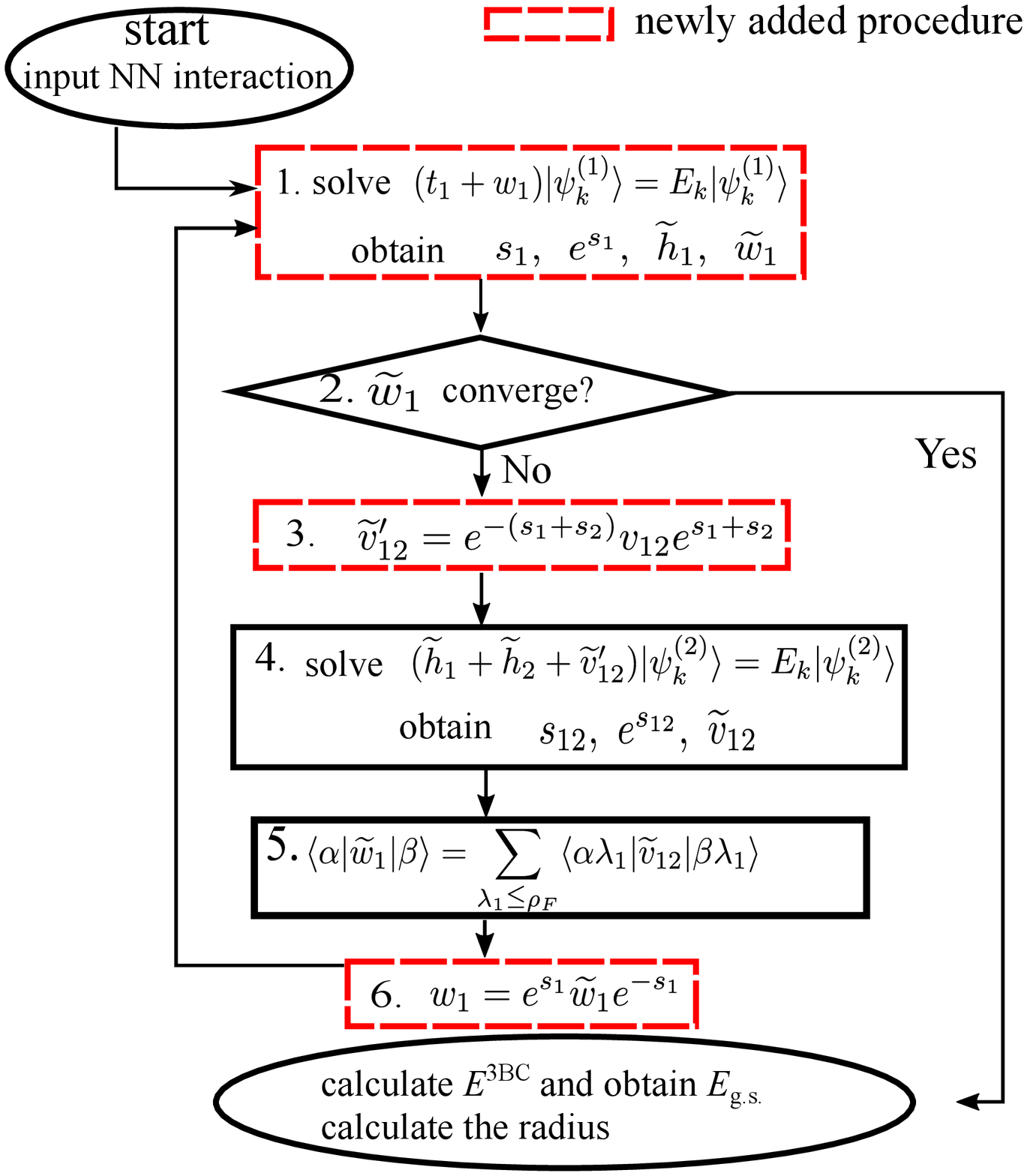}
\caption{\label{flow_chart} (color online) Flow chart of the actual calculation.
The steps surrounded with the dashed line (1. 3. and 6.)
are related to the procedure on the $1p1h$ decoupling newly added this time.
See text for details.}
\end{figure}

In this subsection, we discuss the numerical implementation of the UMOA with the $1p1h$ decoupling examined in the present work.
In this study, we treat only two-body interactions and
do not include any three- and higher many-body ones in the initial Hamiltonian,
Eqs.~(\ref{H_ori}) and (\ref{H_ori2}).
In addition, we keep only the one- and two-body cluster terms and correlation operators.
Thus, two- and many-body auxiliary potentials in Eq.~(\ref{def_W})
 are dropped out and only one-body auxiliary potentials $W=\sum_{i}w_{i}$ are left.
Under these conditions, the flow chart of the actual computation is illustrated schematically in Fig.~\ref{flow_chart}.
Each step of the procedure is also listed as follows:
\begin{enumerate}
 \item The one-body Schr\"odinger equation,
       \begin{equation}
         (t_{1} + w_{1})|\psi^{(1)}_{k}\> = E_{k}|\psi^{(1)}_{k}\>,
       \end{equation}
       is solved and $s_{1}$,
       $e^{s_{1}}$, $\widetilde{h}_{1}$, $\widetilde{t}_{1}$, and
       $\widetilde{w}_{1}$ are obtained by using Eqs. (\ref{Sn}),
       (\ref{exp_omega}), and (\ref{1b_int}), respectively.
 \item The calculation is performed iteratively until
   $\max(|\widetilde{w}^{({\rm new})}_{1} - \widetilde{w}^{({\rm old})}_{1}|) < 10^{-5}$
   is satisfied.
   Here, $\widetilde{w}^{({\rm new})}_{1}$ and $\widetilde{w}^{({\rm old})}_{1}$ are
   the one-body fields obtained at the current and previous iteration steps, respectively.
 \item By using the operator $e^{s_{1}}$ obtained in the step 1, the original
       two-body interaction is transformed as
       \begin{equation}
	\widetilde{v}'_{12} = e^{-(s_{1}+s_{2})}v_{12}e^{s_{1}+s_{2}}.
       \end{equation}
 \item The two-body Schr\"odinger equation,
       \begin{equation}
         (P^{(2)} + Q^{(2)})(\widetilde{h}_{1}+\widetilde{h}_{2} +
         \widetilde{v}'_{12})(P^{(2)} + Q^{(2)})|\psi^{(2)}_{k}\> = E_{k}|\psi^{(2)}_{k}\>,
       \end{equation}
       is solved and $s_{12}$,
       $e^{s_{12}}$, and $\widetilde{v}_{12}$ are obtained by using
       Eqs. (\ref{Sn}), (\ref{exp_omega}), and (\ref{2b_int}), respectively.
 \item By taking the normal ordering with respect to the Fermi level,
       \begin{equation}
         \label{w1_appr}
	\<\alpha|\widetilde{w}_{1}|\beta\> = \sum_{\lambda \le
	 \lambda_{F}} \<\alpha\lambda | \widetilde{v}_{12}|\beta\lambda\>,
       \end{equation}
       we obtain new $\widetilde{w}_{1}$.
 \item Applying the inverse transformation,
       \begin{equation}
	w_{1} = e^{s_{1}} \widetilde{w}_{1} e^{-s_{1}},
       \end{equation}
       we get $w_{1}$ to be
       substituted into the one-body Schr\"odinger equation in the step 1.
\end{enumerate}
After the calculation procedure described above, the correlation operators $s_{1}$ and $s_{12}$ are evaluated.
Using $s_{1}$ and $s_{12}$, we can construct the transformed Hamiltonian
and then obtain the ground-state energy and wave function.
Moreover, we can obtain the other expectation values of observables with the same transformed
 operators as discussed in Sec.~\ref{sec:ob}.
This is one of the advantages about effective operators in the UMOA.

\subsection{Ground-state energy and radius \label{sec:ob}}
\begin{figure}[t]
\subfigure[]{
\includegraphics[clip,width=1.5in]{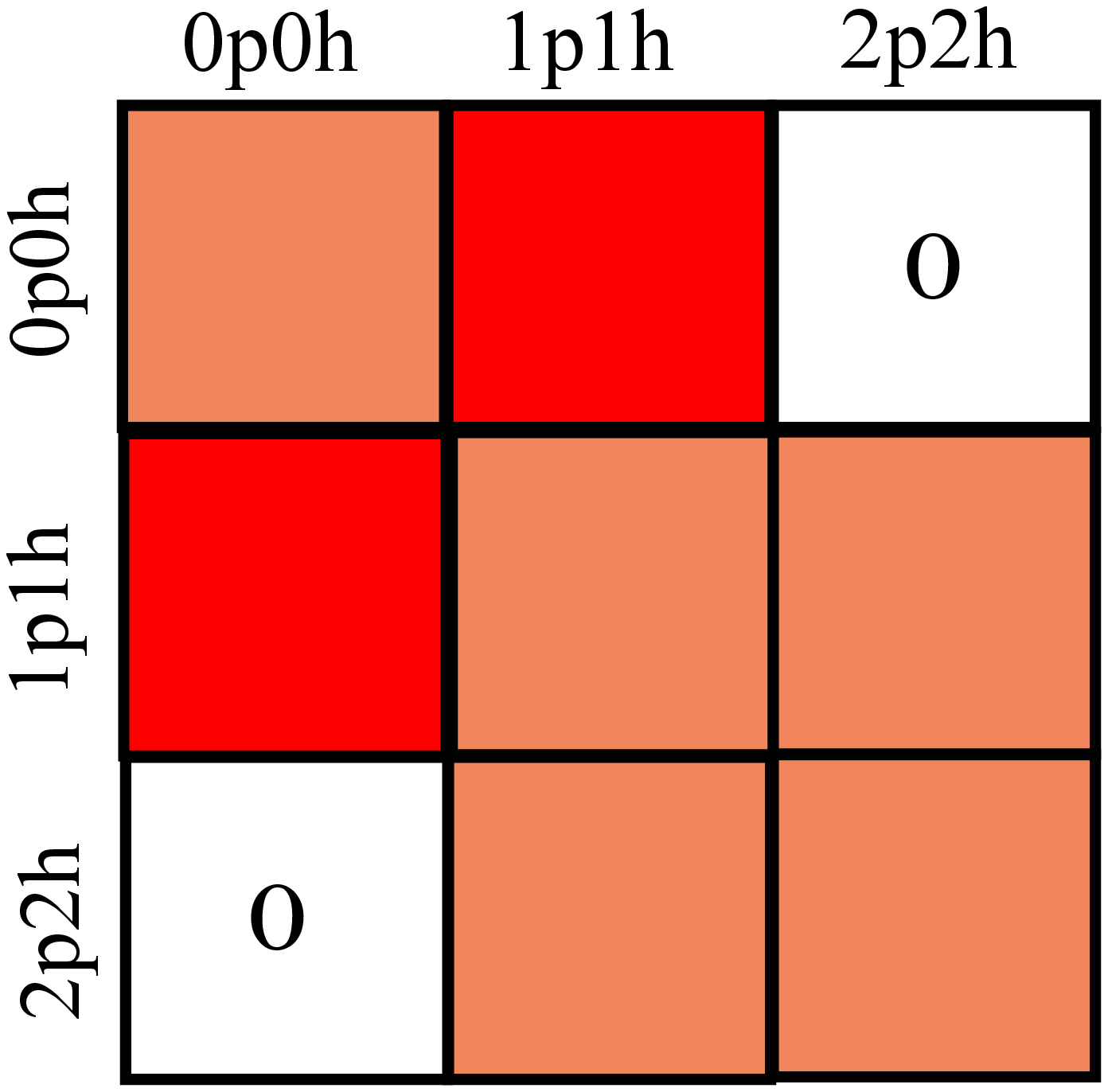}
\label{wo_1b}}
\subfigure[]{
\includegraphics[clip,width=1.5in]{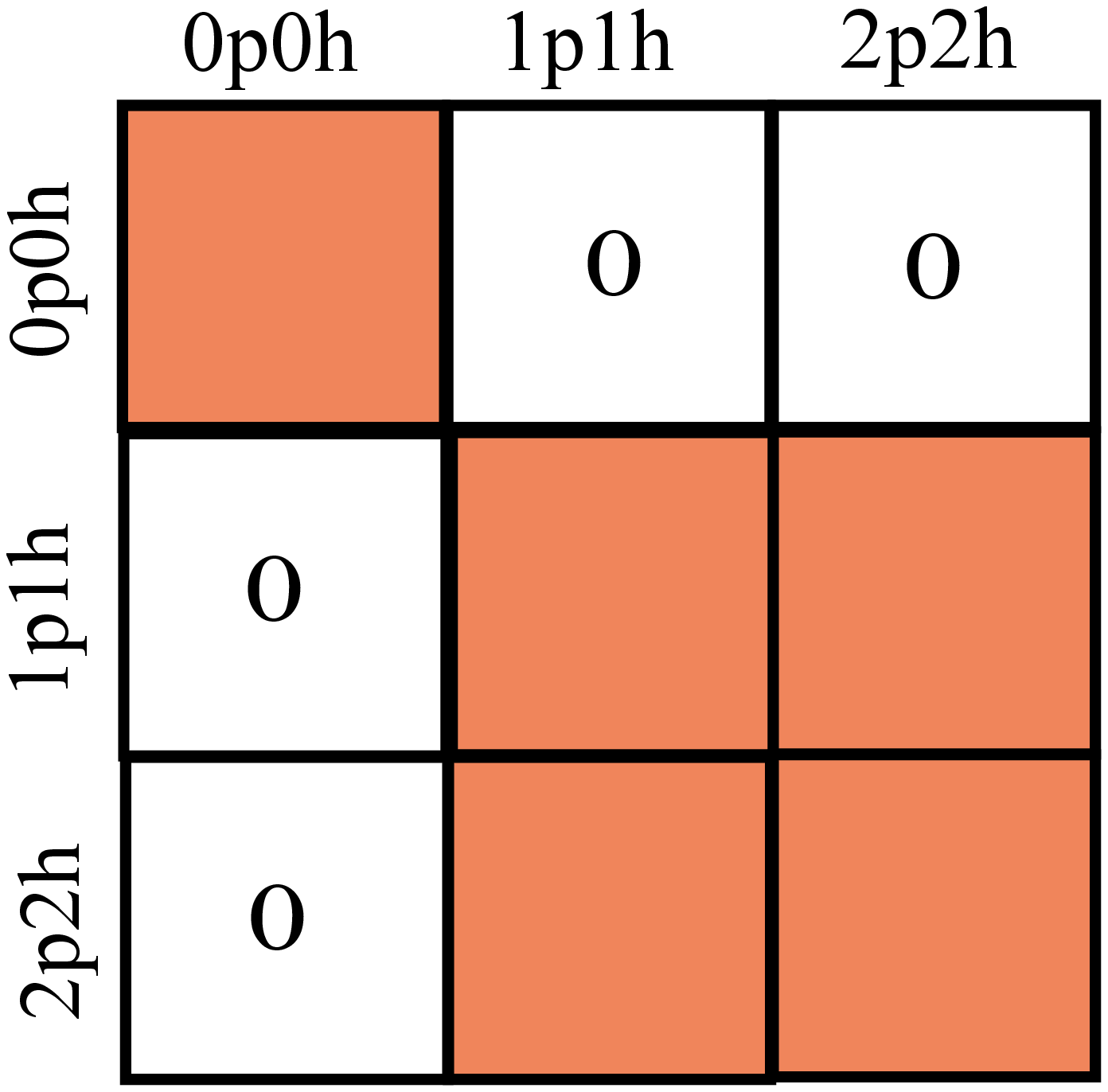}
\label{w_1b}}
\caption{\label{Tran_H} (color online) Schematic representations of the transformed Hamiltonians
  without (left; (a)) and with (right; (b)) the one-body correlation operator.}
\end{figure}

Here, we present how to calculate the ground-state energy and root-mean square radius in the UMOA.
 After the transformation discussed in Sec. \ref{sec:cal}, we construct the
 transformed Hamiltonian up to the two-body parts with Eqs. (\ref{1b_h}),
 (\ref{2b_h}), (\ref{1b_int}), (\ref{2b_int}), and (\ref{w1_appr}).
Fig. \ref{Tran_H} shows the schematic representations of the transformed Hamiltonians with and without the one-body correlation operator.
In the previous publications \cite{Miyagi:2014,Miyagi:2015}, we have employed
 the Hamiltonian without the $1p1h$ decoupling ($S^{(1)}=0$) as shown schematically in Fig. \ref{wo_1b} to evaluate the ground-state
energy and radius.
Instead, we diagonalized the transformed Hamiltonian in the $0p0h$ and $1p1h$ space
 to take into account the effect of off-diagonal components between the $0p0h$ and $1p1h$ states.
While, in the present formulation, the matrix elements between the $0p0h$ and $1p1h$ states vanish from the beginning,
because of the $1p1h$ decoupling by the one-body correlation operator $S^{(1)}$ as shown in Fig~\ref{w_1b}.
The ground-state energy is dictated with the transformed Hamiltonian and reference state as
\begin{equation}
 \label{egs}
  E_{\text{g.s.}} = \<\Psi|H|\Psi\> =
(\<\Psi|U) U^{\dag}HU (U^{\dag}|\Psi\>)= \<\Phi|\widetilde{H}|\Phi\>.
\end{equation}
The rightmost term in Eq. (\ref{egs}) corresponds to the zero-body term after
taking the normal ordering with respect to the reference state $|\Phi\>$.
It is approximated up to the third order of expansion as
\begin{align}
 \label{egs_2}
 E_{\rm g.s.} &\approx \sum_{\lambda \le \rho_{F}} \<\lambda|
 \widetilde{h}_{1}|\lambda\> +
 \frac{1}{2!}\sum_{\lambda\mu\le\rho_{F}}\<\lambda\mu|\widetilde{v}_{12}|\lambda\mu\>
 \notag \\
&\hspace{3em} -
 \sum_{\lambda\le\rho_{F}}\<\lambda|\widetilde{w}_{1}|\lambda\> + E^{{\rm 3BC}} \\
 \label{egs_3}
 &\hspace{-2em}= \sum_{\lambda \le \rho_{F}} \<\lambda|
 \widetilde{t}_{1}|\lambda\> +
 \frac{1}{2!}\sum_{\lambda\mu\le\rho_{F}}\<\lambda\mu|\widetilde{v}_{12}|\lambda\mu\>
  + E^{{\rm 3BC}},
\end{align}
where $\widetilde{t}_{1} (= e^{-s_{1}} t_{1}e^{s_{1}})$ is the one-body transformed kinetic energy term.
The three-body cluster term $E^{{\rm 3BC}}$ is evaluated {\it a posteriori} as
\begin{equation}
E^{\text{3BC}}= \frac{1}{3!} \sum_{\lambda\mu\nu \le \rho_{F}}
 \<\lambda\mu\nu|\widetilde{v}_{123}|\lambda\mu\nu\>,
\end{equation}
 which is taken into account through order $S^{(2)2}$ \cite{Suzuki:1994}.
From our experience, this approximation is rather good.
The convergence with respect to the cluster expansion will be discussed in Sec.\ref{sec:res}.

One of the advantages in the UMOA is the usage of the same transformed
 operators for energies to obtain the other observables.
 Although, in principle, one can also use the transformed operator in the CCM,
one has to solve the coupled-cluster equation for both of
 left and right eigenvectors to construct the transformation operator due to the non-Hermiticity.
In practical applications, the Hellmann-Feynman theorem is often applied
to obtain the observables other than energies, as found in the recent studies (see e.g., Ref~\cite{Hagen:2014}).
On the other hand, the calculation of observables is rather simple in the UMOA.
For an explicit demonstration, we show how to calculate the radius in the following.
Let $r^{2}$ be the squared point-nucleon radius operator defined as
\begin{equation}
  r^{2} = \frac{1}{A}\sum_{i}({\mathbf r}_{i} - {\mathbf R}_{{\rm c.m.}})^{2}
\end{equation}
with the coordinate vector of $i$-th nucleon ${\mathbf r}_{i}$ and of
 the center-of-mass ${\mathbf R}_{{\rm c.m.}}$.
Similarly to the evaluation of ground-state energy,
the squared radius operator $r^{2}$ is decomposed into the one- and two-body operators,
\begin{equation}
  r^{2}=r^{2(1)}+r^{2(2)},
\end{equation}
where $r^{2(1)}$ and $r^{2(2)}$ are the one- and two-body parts, respectively, as
\begin{align}
  r^{2(1)} &= \frac{1}{A} \left(1 - \frac{1}{A} \right)
  \sum_{i=1}^{A} \mathbf{r}^{2}_{i} = \sum_{i = 1}^{A} r_{i}^{2}, \\
 r^{2(2)}&=  - 2\frac{1}{A^{2}} \sum_{i<j}^{A}
  \mathbf{r}_{i} \cdot \mathbf{r}_{j} = \sum_{i<j}^{A} r_{ij}^{2}.
\end{align}
With the aid of the unitary operator $U$, the expectation value
can be expressed by the transformed operator $\widetilde{r}^{2}$ and the reference state $|\Phi\>$ as
\begin{equation}
  \<\Psi|r^{2}|\Psi\> = \<\Phi|\widetilde{r}^{2}|\Phi\>
\end{equation}
with the transformed radius operator $\widetilde{r}^{2}$
\begin{equation}
  \widetilde{r}^{2} = U^{\dag}r^{2} U.
\end{equation}
Then, we carry out the cluster expansion of the transformed operator $\widetilde{r}^{2}$,
\begin{equation}
  \widetilde{r}^{2}=  \widetilde{r}^{2(1)}+
  \widetilde{r}^{2(2)} + \cdots.
\end{equation}
Here, the one- and two-body cluster terms are generated as
\begin{equation}
  \widetilde{r}^{2(1)}= \sum_{i}\widetilde{r}_{i}^{2}, \qquad
  \widetilde{r}^{2(2)}= \sum_{i<j}\widetilde{r}_{ij}^{2}
\end{equation}
with
\begin{align}
  \widetilde{r}_{1}^{2} &= e^{-s_{1}}r_{1}^{2} e^{s_{1}}, \\
  \widetilde{r}_{12}^{2} &= e^{-s_{12}}e^{-(s_{1}+s_{2})} (r_{1}^{2}+r_{2}^{2}+r_{12}^{2})
  e^{s_{1}+s_{2}}e^{s_{12}} \notag \\
  & \hspace{15em}- (\widetilde{r}_{1}^{2}+\widetilde{r}_{2}^{2}).
\end{align}
The expectation value is equal to the normal-ordered zero-body term,
\begin{align}
  \<\Phi|\widetilde{r}^{2}|\Phi\>
&= \sum_{\lambda \le \rho_{F}}
  \<\lambda|\widetilde{r}^{2(1)}|\lambda\> +
 \frac{1}{2}\sum_{\lambda\mu\le\rho_{F}}
 \<\lambda\mu|\widetilde{r}^{2(2)}|\lambda\mu\> + \cdots \nonumber \\
 \label{EffOp2}
  & \approx \widetilde{r}^{2 ({\rm 1BC})} + \widetilde{r}^{2({\rm 2BC})}.
\end{align}
Here, $\widetilde{r}^{2 ({\rm 1BC})}$ and $\widetilde{r}^{2 ({\rm 2BC})}$
 are the contributions of one- and two-body transformed radius operators, respectively.
In actual calculations, we truncate the cluster expansion up to the second order to evaluate the radius as in Eq.~(\ref{EffOp2}).

\section{\label{sec:res}Results and discussion}
In the earlier UMOA studies, the ground-state properties of $^{16}$O, $^{40}$Ca, and $^{56}$Ni have been mainly discussed with
realistic $NN$ interactions \cite{Suzuki:1994, Fujii:2004, Fujii:2009, Miyagi:2015, Miyagi:2014}.
Only the decoupling of the $2p2h$ excitations has been considered, i.e.,  $U = e^{S^{(2)}}$ in Eq. (\ref{uni_op}).
The calculated results, especially for radii, strongly depend on the $\hbar\omega$ values
and are difficult to judge the reliability from {\it ab-initio} point of view.
Following the success to reduce the $\hbar\omega$-dependence in the CCM \cite{Kohno:2012},
 we naturally extend the formalism and introduce the one-body correlation
 operator $S^{(1)}$ to the UMOA in the present study.
As the validation of the effectiveness, we show the numerical results of ground-state energy
and point-nucleon radius of $^{4}$He
 with the transformation $U = e^{S^{(1)}}e^{S^{(2)}}$.
 Then, we discuss the role of the one-body correlation operator $S^{(1)}$ in the UMOA to some extent.

The choice of the initial Hamiltonian for numerical calculations is one of the important issues.
Nowadays, the sophisticated nucleon-nucleon ($NN$) interactions have been developed and
 reproduce the $NN$ scattering phase shift data with the high precision,
 as well as the deuteron properties such as the AV18~\cite{Wiringa:1995}, CD-Bonn~\cite{Machleidt:2001},
  and chiral EFT N$^{3}$LO interactions~\cite{Entem:2003}.
It is, however, difficult to apply directly such bare $NN$ interactions to our calculations,
 because of the strong coupling between low- and high-momentum regions.
To get rid of this computational difficulty, we have applied on the earlier UMOA the
 effective interactions derived with some approximations
which hamper reliable estimations of the uncertainty on calculated results
 (see e.g. Refs. \cite{Suzuki:1994,Fujii:2004}).

In the present work, we are interested in confirming the applicability
 of the UMOA through the comparison to the other {\it ab initio} results.
It is preferable to reduce the uncertainties coming from the effective interactions
employed in the earlier UMOA as much as possible.
Therefore, we omit such process in the earlier UMOA by using sophisticated softened interactions.
For this purpose, we mainly use two types of $NN$ interactions.
One is the low-momentum interaction
 $V_{{\rm low \ k}}$ derived from the AV18 $NN$ interaction~\cite{Wiringa:1995} with the sharp cutoff $\Lambda = 1.9$ fm$^{-1}$.
The other is the SRG-transformed chiral EFT N$^{3}$LO $NN$ interaction~\cite{Entem:2003} with the cutoff
  $\lambda_{\rm SRG} = 2.0$ fm $^{-1}$
 to compare with the recent {\em ab initio} calculation results~\cite{Nogga:2004, Hagen:2007, Roth:2011}.
 As the qualitative aspect between these two interactions is similar enough to discuss the role of $S^{(1)}$ in the UMOA,
 we mainly show the results with the low-momentum interaction $V_{{\rm low \ k}}$ in the following discussion.
 Note that our results are not comparable directly to the experimental data
due to the missing genuine three-body and induced many-body interactions
which cannot be treated in the $V_{\rm low \ k}$ formalism.
We can, however, compare our present UMOA results with the other {\it ab initio} ones
obtained by using the same $NN$ interactions.


\begin{figure*}[t!]
\includegraphics[clip,width=15cm, clip]{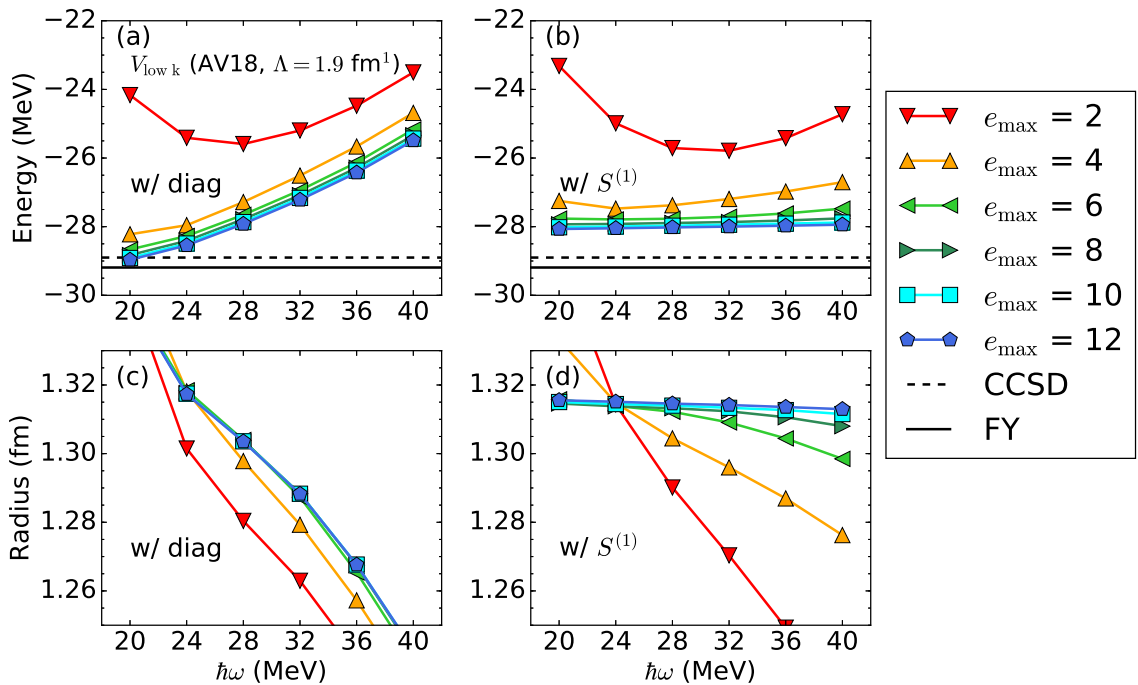}
\caption{\label{4He_conv_vlowk} (color online) Ground-state energies (upper panels; (a) and (b)) and
  point-nucleon root-mean-square matter radii (lower panels; (b) and (d)) of the $^{4}$He nucleus
 as a functions of the HO energy $\hbar\omega$.
 In (a) and (c), one-body correlations are approximately taken into account through the
 diagonalization in 0$p$0$h$ and 1$p$1$h$
 space as done in earlier works~\cite{Fujii:2004, Fujii:2009, Miyagi:2014, Miyagi:2015}.
 In (b) and (d), one-body correlations are explicitly included.
Note that the results from ``FY'' (solid line) and ``CCSD'' are taken from
Refs.~\cite{Nogga:2004} and~\cite{Hagen:2007}, respectively.}
\end{figure*}

\begin{figure}[t!]
\includegraphics[clip,width=7cm, clip]{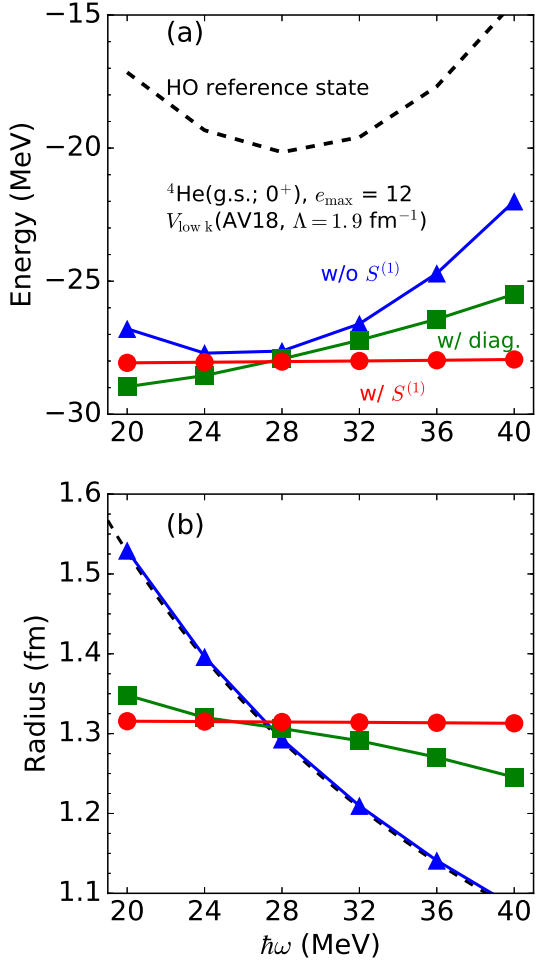}
\caption{\label{4He_emax_1b_vlowk} (color online) Ground-state energies (upper panel) and
 point-nucleon root-mean-square matter radii (lower panel) of the $^{4}$He nucleus
 as a functions of the HO energy $\hbar\omega$.
Circles and triangles are given with the Hamiltonian in Fig.~\ref{w_1b} and
 Fig.~\ref{wo_1b}, respectively.
Squares are obtained after the diagonalization
 of the Hamiltonian in Fig.~\ref{wo_1b} in the $0p0h$ and $1p1h$ space.
The dashed curves are obtained with the $0p0h$ HO reference state.}
\end{figure}

\begin{figure}[t!]
\includegraphics[clip,height=19.6cm, clip]{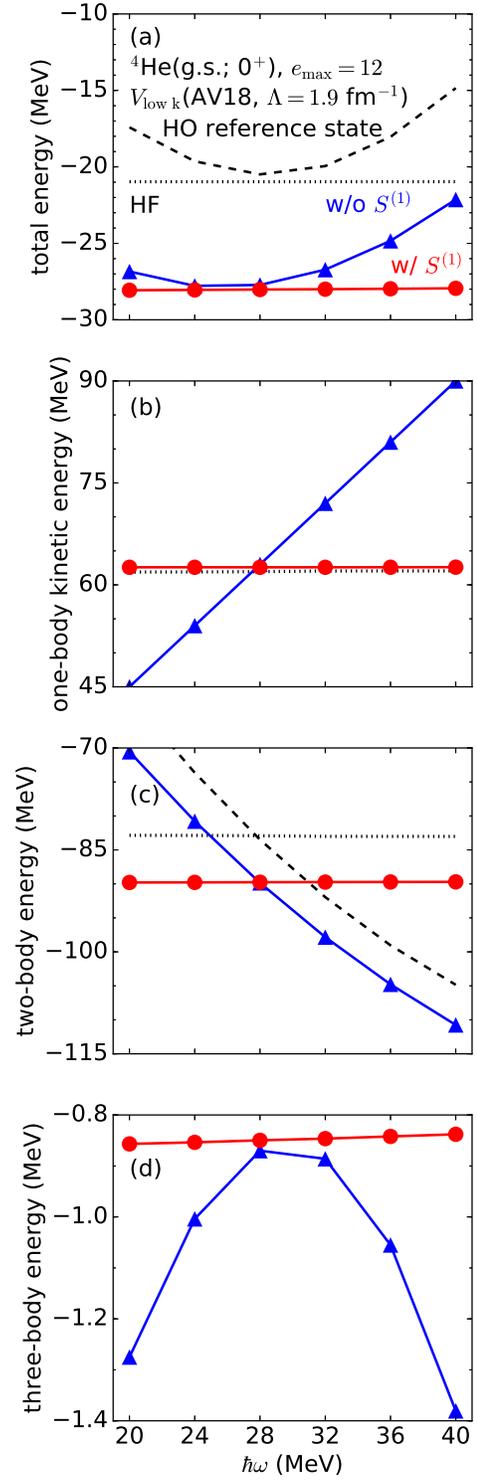}
\caption{\label{T_U} (color online) The total (a), one-body (b), two-body (c), and three-body energies (d)
for $^{4}$He as functions of $\hbar\omega$ from the top to the bottom.
The $NN$ interaction is the $V_{\rm low \ k}$ interaction
   derived from AV18 interaction~\cite{Wiringa:1995} at $\Lambda = 1.9$ fm$^{-1}$.
Circles (triangles) correspond to the results with (without) $S^{(1)}$.
The dashed and dotted lines are for the HO reference and HF states, respectively.}
\end{figure}

\begin{figure}[t]
\includegraphics[clip,width=\linewidth, clip]{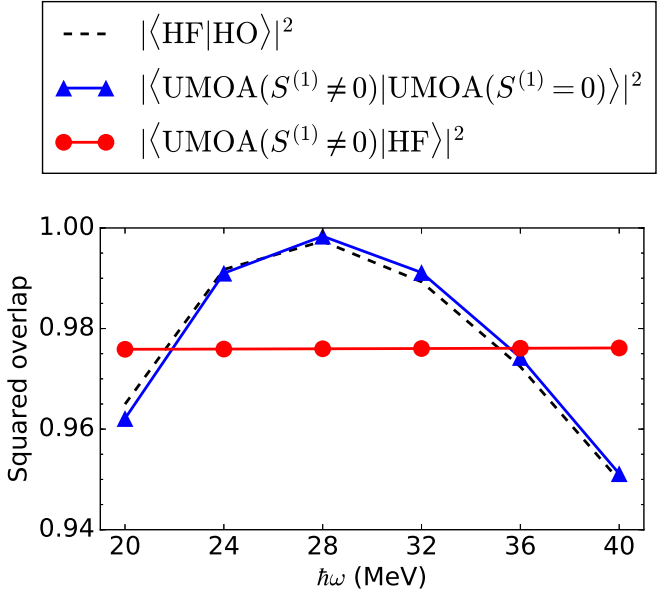}
\caption{\label{ovlap} (color online) The squared overlaps between the ground and
 reference states as functions of $\hbar\omega$.
 The employed $NN$ interaction is the $V_{\rm low \ k}$ interaction
   derived from AV18 interaction at $\Lambda = 1.9$ fm$^{-1}$.
 Circles (triangles) are obtained with the UMOA with $S^{(1)}$ and HF (UMOA without $S^{(1)}$) states.
 The dashed line is given with the HF and HO reference states.}
\end{figure}

\begin{figure}[t!]
\includegraphics[clip,width=\linewidth, clip]{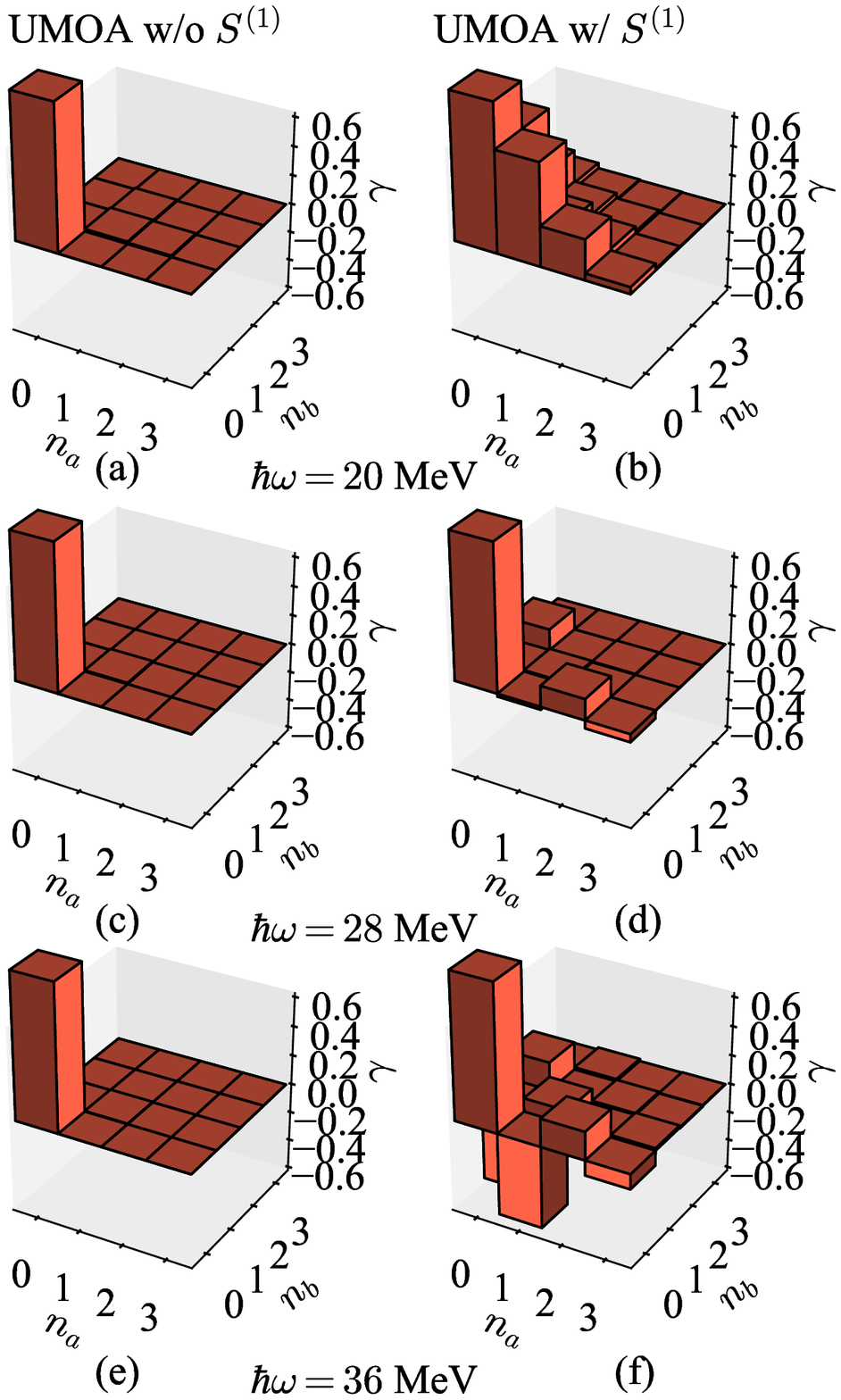}
\caption{\label{fig:density_matrix}(color online)
  One-body density matrix $\gamma$ of $s1/2$ orbitals for $^{4}$He calculated without (left; (a), (c), and (e)) and with (right; (b), (d), and (f))
 the one-body correlation operator.
 The top, middle, and bottom panels are calculated at $e_{\rm max}=12$
 and $\hbar\omega =$20, 28, and 36 MeV, respectively.}
\end{figure}
\subsection{\label{Sec:Vlowk}Role of the one-body correlation operator}

Before discussing the importance of one-body correlations in the UMOA,
Fig.~\ref{4He_conv_vlowk} summarizes the dependence of the model-space size
on the energy and point-nucleon radius of the $^4$He ground state in the former and current UMOA.
In the left panels, one-body correlations are approximately taken into account
through the diagonalization in the $0p0h$ and $1p1h$ space as done in earlier works \cite{Fujii:2004, Fujii:2009, Miyagi:2014, Miyagi:2015}.
While, in the right panels, one-body correlations are explicitly included by the one-body correlation operator as in the present formalism.
As seen in the figure, the convergence with respect to the size of the model space can be almost achieved around $e_{\max} = 12$
both in the earlier (in the left panels) and current (in the right panels) formulations.
Interestingly, in the earlier formalism (shown in the left panels),
we can observe the artificial $\hbar\omega$-dependence even after the convergence for the model-space size is achieved.
It is our motivation behind the employment of the one-body correlation operator in the UMOA
so as to eliminate this artificial dependence.
On the other hand, in the current formalism (shown in the right panels),
the results show typical behavior for the $\hbar\omega$ and $e_{\max}$ parameters
found in usual {\it ab initio} calculations, i.e., the $\hbar\omega$-dependence reduces as the model-space size increases.
Now we can obtain the converged results by the UMOA calculations with sufficiently large model space differently than before.
In the figure, the FY~\cite{Nogga:2004} and CCSD~\cite{Hagen:2007} results for the ground-state energy with the same $NN$ interaction are also shown for a comparison purpose.

After confirming the convergence of the results with respect to the parameters, $\hbar\omega$ and $e_{\rm max}$,
Fig.~\ref{4He_emax_1b_vlowk} compares the ground-state energies (upper panel)
and point-nucleon root-mean-square radii (lower panel) of
 $^{4}$He as functions of $\hbar\omega$ in the earlier and current UMOA formalisms
with and without the one-body correlation operator $S^{(1)}$.
Note that we show the results only in $e_{\rm max} = 12$, because
 the calculated results almost converge at $e_{\max} = 12$ as found in Fig.~\ref{4He_conv_vlowk}.
For example, the difference between our results for energies
  at $e_{\max} = 12$ and $e_{\max} = 14$ is order of 10 keV.
 It is sufficient for the present purpose of discussion.
In the figure, the results without the one-body correlation operator $S^{(1)}$ (blue triangles) are almost parallel
with the HO $0p0h$ results (black dashed line) both for the ground-state energies and point-nucleon radii.
It implies that the operator $S^{(2)}$ does not change the trend of the $\hbar\omega$-dependence originated from the HO $0p0h$ reference state.
After the diagonalization in the $0p0h$ and $1p1h$ space (green squares)
as employed in the earlier framework~\cite{Fujii:2004, Fujii:2009, Miyagi:2014, Miyagi:2015},
 the $\hbar\omega$-dependence on the results is slightly reduced compared to the UMOA without $S^{(1)}$ (blue triangles).
In contrast, the situation drastically changes, if the transformation operator $U$ is constructed
by the one- and two-body correlation operators as in the current formalism.
The results from the UMOA with $S^{(1)}$ (red circles) are practically $\hbar\omega$-independent both for the energy and radius.
This is the main result in this paper showing that the one-body correlation operator functions well in the UMOA
to reduce the $\hbar\omega$-dependence, which is needed to obtain reliable converged results.
In the present framework, we do not have to care about the choice of $\hbar\omega$, if the calculations
 are done in sufficiently large model spaces.

There are additional remarks on the optimum $\hbar\omega$ value in the UMOA.
In the CCM \cite{Kohno:2012}, the choice of the $\hbar\omega$ values is discussed by comparing the
 results with and without the one-body correlation operator.
 In Ref. \cite{Kohno:2012}, the authors concluded that the results (calculated without the one-body correlation operator)
  at $\hbar\omega$ minimizing the ground-state energy agree with the $\hbar\omega$-independent
  results (calculated with the one-body correlation operator).
This statement is confirmed by the present UMOA calculations.
Our results for energies with (circles) and without (triangles) the $S^{(1)}$ operator
 are very close to each other around the optimum $\hbar\omega$ value of $\hbar\omega = 28$ MeV in Fig~\ref{4He_emax_1b_vlowk}.
There is, however, a word of caution about the choice of the $\hbar\omega$ value for the results after the diagonalization
in the earlier UMOA.
In our previous investigations \cite{Miyagi:2014,Miyagi:2015}, the ground-state energies and
 charge radii of doubly magic nuclei have been calculated in the UMOA without $S^{(1)}$.
Then, the results converge with respect to the model-space size, while
 the $\hbar\omega$-dependence of the results, especially for charge radii, is not negligible.
To determine accurately the radii, we have taken the results after the diagonalization at
 $\hbar\omega$ minimizing the ground-state energy,
corresponding to the square symbol at $\hbar\omega=20$ MeV in Fig.~\ref{4He_emax_1b_vlowk}.
Since we do not confirm the agreement of the results with the diagonalization (squares) and
 with $S^{(1)}$ (circles) at $\hbar\omega=20$ MeV in Fig.~\ref{4He_emax_1b_vlowk},
 such an estimation would not be valid after the diagonalization.
In the present framework, however, we do not have to care about the choice of
 $\hbar\omega$, because of the weak $\hbar\omega$-dependence of the results.

After glancing over the effect of the one-body correlation operator,
we next discuss how the $\hbar\omega$-dependence of the results is reduced
 by introducing the $S^{(1)}$ operator.
To look closer the role of the one-body correlation operator, it is convenient to
 decompose the expectation value into the contributions from each cluster term.
In the case of the ground-state energy, it can be decomposed into energies from
the one-body kinetic term $\widetilde{t}_{1}$, two-body cluster term
 $\widetilde{v}_{12}$, and three-body cluster term $\widetilde{v}_{123}$.
Fig.~\ref{T_U} describes the energies from each decomposed cluster term.
The total, one-body, two-body, and three-body energies
of the $^{4}$He ground state are shown as functions of $\hbar\omega$ from the top to the bottom in the figure.
As a reference, the HO (dashed line) and HF (dotted) reference states are also drawn.
From the figure, one can find that the contribution from the three-body cluster term
is smaller by two orders of magnitude than the one- and two-body cluster terms.
The cluster expansion in the UMOA converges in the case of the $^4$He ground-state energy.
Also the $\hbar\omega$-dependence of the energy expectation values is reduced at each order of the cluster expansion.

Now, let us focus on the one-body kinetic energy part (the second panel from the top in Fig.~\ref{T_U}).
When we ignore $S^{(1)}$ from the beginning ($S^{(1)} = 0$),
 $\<t_{1}\>$ is nothing but the sum of the diagonal component of the matrix $\<a|t_{1}|a\>$.
Here, $\<X \>$ means the expectation value of an operator $X$
 with respect to the HO $0p0h$ reference state $|{\rm HO}\>$.
In the case of the $^4$He ground state,
one can easily find that $\<t_{1}\>$ (denoted by the blue triangles in the second panel from the top in Fig.~\ref{T_U})
 is proportional to $\hbar\omega$ as
\begin{align}
\<t_{1}\> &= \left( 1 - \frac{1}{A}
    \right)\sum_{a \le \rho_{F}} \left( 2n_{a} + l_{a} +
    \frac{3}{2}\right) \frac{\hbar\omega}{2} \notag \\
    \label{eq:one_ho}
    & = \frac{9}{4} \hbar\omega,
\end{align}
with respect to the HO reference state (0s1/2)$^{4}$.
On the other hand, when $S^{(1)}$ is introduced,
 the off-diagonal component of the original matrix $\<a|t_{1}|b\>$ $(a \neq b)$
contributes to $\<\widetilde{t}_{1}\> = \< e^{-s_{1}}t_{1}e^{s_{1}}\>$.
 As a result, $\< \widetilde{t}_{1} \>$ becomes practically $\hbar\omega$-independent
(as denoted by the red circles in the figure).
Also, these are quite close to the HF
 kinetic energy $\<{\rm HF} | t_{1}| {\rm HF}\>$ (black dotted curve in the figure).
Thus, the $S^{(1)}$ operator acts as the generator of the transformation from
$|{\rm HO}\>$ to $|{\rm HF}\>$ in the UMOA.

Almost the same discussion can be done for the expectation value
 of the two-body cluster term.
The dashed line in the third panel from the top in Fig.~\ref{T_U} corresponds to
\begin{equation}
 \< v_{12} \> = \frac{1}{2}\sum_{a,b \le \rho_{F}}\<ab|v_{12}|ab\>,
\end{equation}
which is the sum of the diagonal components of the original two-body matrix elements.
As seen in the figure, it shows the monotonic dependence on $\hbar\omega$.
This tendency can be understood by assuming a simple $S$-wave two-nucleon potential model
 such as the contact interaction regularized by the Gaussian with a cutoff momentum $\Lambda_{\delta}$:
 \begin{equation}
   V_{\delta} = \left\{
     \begin{array}{c}
       C_{^{1}S_{0}} \exp\left( -
       \frac{q^{2} + q'^{2}}{\Lambda_{\delta}^{2}}
     \right) \ {\rm for} \ ^{1}S_{0}\ {\rm channel}, \\
     C_{^{3}S_{1}} \exp\left( -
     \frac{q^{2} + q'^{2}}{\Lambda_{\delta}^{2}}
   \right) \ {\rm for} \ ^{3}S_{1} \ {\rm channel}.
 \end{array}
 \right.
  \label{eq:gaus}
 \end{equation}
Here, $q$ and $q'$ are the magnitudes of relative momenta
 for the initial and final states, respectively.
The $C_{^{1}S_{0}}$ and $C_{^{3}S_{1}}$ are the low-energy constants for
 $^{1}S_{0}$ and $^{3}S_{1}$ channels, respectively.
Note that only the $S$-wave potentials are enough for the discussion about the $(0s1/2)^{4}$ single-particle configuration.
As discussed in Appendix~\ref{sec:toymodel}, one can obtain
\begin{equation}
  \<V_{\delta}\> = 3 \sqrt{\frac{\pi}{2}} (C_{^{1}S_{0}} + C_{^{3}S_{1}})
  \left(
    \frac{\sqrt{\hbar m\omega} \Lambda_{\delta}^{2}}
    {m\omega + \hbar\Lambda_{\delta}^{2}}
  \right)^{3}.
  \label{eq:vcnt}
\end{equation}
The $\hbar\omega$-dependence is given by the derivative of  $\<V_{\delta}\>$ with respect to $\omega$.
As shown in Eq.~(\ref{eq:deri}), the derivative is always negative
 in our $\hbar\omega$ range from 20 to 40 MeV.
While the HF result (dotted curve) shows rather weaker $\hbar\omega$-dependence.
Similar behavior can be found in the UMOA with and without $S^{(1)}$ (circles and triangles, respectively).
Accordingly, the $\hbar\omega$-dependence is reduced by the effect of $S^{(1)}$.

Similarly to the one- and two-body cluster terms, the $\hbar\omega$-dependence on
 the three-body cluster term is also reduced as shown in the bottom panel of Fig.~\ref{T_U}.
From these considerations, the inclusion of the $S^{(1)}$ operator
mitigates considerably the $\hbar\omega$-dependence of the energy expectation value
at each order of the cluster expansion.
Moreover, the relations between the UMOA with and without $S^{(1)}$
resemble those between the results with the HF and HO $0p0h$ states.
Note that, if we ignore $S^{(2)}$ from the beginning,
calculated ground-state energies numerically coincide with the HF ground-state energies within a few keV level.
Thus, a constant shift in the two-body energy comes from the $S^{(2)}$ contribution.

To examine more directly the role of $S^{(1)}$,
 the overlap of wave functions is also investigated.
In Fig.~\ref{ovlap}, we show three squared overlaps
obtained with the HF and HO reference states (black dashed line),
the UMOA with and without $S^{(1)}$ (blue triangles),
 and the UMOA with $S^{(1)}$ and HF states (red circles).
Note that the UMOA results are obtained at $e_{\max} = 12$.
The squared overlap $|\<{\rm HO}|{\rm HF}\>|^{2}$ (dashed line) indicates the effect of the optimization of
 single-particle basis states.
In Fig.~\ref{ovlap}, the squared overlaps $|\<{\rm HO}|{\rm HF}\>|^{2}$ (black dashed line) and
$|\< {\rm UMOA} (S^{(1)} \neq 0)|{\rm UMOA} (S^{(1)} = 0) \>|^{2}$
(blue triangles) behave in a similar way.
Therefore, the role of $S^{(1)}$ is to optimize the single-particle basis states,
 as expected from the above discussion of the ground-state energy.
As a check, we confirm that the overlap $|\< {\rm UMOA}(S^{(1)} \neq 0)|{\rm HF}\>|^{2}$ (red circles)
 does not depend on the $\hbar\omega$ values.

Further insight can be acquired by looking into the one-body density matrix $\gamma$.
The derivation of the one-body density matrix in the UMOA is shown in Appendix~\ref{Sec:density}.
Using Eq.~(\ref{tr-exp}), we have $\<\widetilde{t}_{1}\>_{\rm L.O.} = {\rm Tr}(\gamma t_{1})$ at the leading order of cluster expansion.
Fig.~\ref{fig:density_matrix} shows the density matrices for the $s1/2$ orbital with the $\hbar\omega$ values
ranging from 20 to 36 MeV at $e_{\max} = 12$.
For visibility of the figure, we only show the density matrices up to the components of max($n_a$, $n_b$) $= 3$.
When the one-body correlation operator is switched off ($S^{(1)} = 0$), only the component of the density matrix $\gamma_{0s1/2,0s1/2}$ is dominant.
Then, as shown in Eq.~(\ref{eq:one_ho}),
$\<t_{1}\>$ linearly increases as a function of $\hbar\omega$.
In contrast, the off-diagonal elements, especially $\gamma_{0s1/2,1s1/2}$,
 can have large values, when the one-body correlation operator is turned on ($S^{(1)} \neq 0$).
At $\hbar\omega=20$ MeV, $\gamma_{0s1/2, 1s1/2}$ is positive and increases $\<\widetilde{t}_{1}\>$.
Note that the off-diagonal component of $\<a|t_{1}|b\>$ is always positive.
On the other hand, $\gamma_{0s1/2, 1s1/2}$ is negative and decreases $\<\widetilde{t}_{1}\>$ at
 $\hbar\omega=36$ MeV.
 The contribution from the off-diagonal components to $\<\widetilde{t}_{1}\>$ balances around $\hbar\omega=28$ MeV,
  because $\gamma_{0s1/2,1s1/2}$ is almost zero there.
This finding is consistent with the behavior of one-body energies shown in Fig.~\ref{T_U}.

Next, we discuss the effects of the one-body correlation operator on the point-nucleon radius.
Since the radius operator is dominated by the one-body term,
the one-body density matrix concerns more directly the reduction of the
 $\hbar\omega$-dependence rather than the ground-state energy where the two-body correlations are dominant.
As shown in the lower panel in Fig.~\ref{4He_emax_1b_vlowk},
the point-nucleon radius of the $^{4}$He nucleus
with respect to the HO reference state (0s1/2)$^{4}$ can read
\begin{align}
  \sqrt{\< r^{2}_{1}\>}  &=
  \sqrt{\sum_{a}\frac{1}{A}\left(1 - \frac{1}{A}\right) \frac{(\hbar c)^{2}}
  {mc^{2} \hbar\omega} \left(2n_{a} + l_{a} + \frac{3}{2}\right)} \notag \\
  \label{r1}
  &\simeq 6.8 \left( \hbar\omega \right)^{-1/2} ({\rm MeV})^{1/2} {\rm fm}.
\end{align}
From the observation that the off-diagonal components of the kinetic and radius one-body operators contribute oppositely,
 we can expect that the radius decreases (increases) in smaller (larger) $\hbar\omega$ region across $\hbar\omega \simeq 28$ MeV
 compared to the HO reference state Eq.~(\ref{r1}).
This is consistent with the radii shown in the lower panel of Fig.~\ref{4He_emax_1b_vlowk}.

\begin{figure}[t!]
\includegraphics[clip,width=\linewidth, clip]{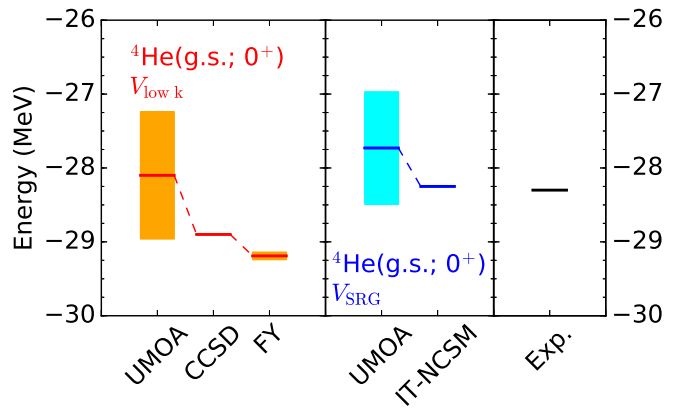}
\caption{\label{4He_final} (color online) Comparisons for the ground-state
 energies of $^{4}$He in the UMOA and in the other {\it ab initio} calculations.
 The displayed UMOA results are obtained at $e_{\max} = 14$ and $\hbar\omega$ = 20 MeV.
 The error bands of the UMOA energies are estimated from the size of the three-body cluster term corrections.
 The CCSD, FY, IT-NCSM, and experimental results are taken from Refs.~\cite{Hagen:2007},
 \cite{Nogga:2004, Hagen:2007}, \cite{Roth:2011}, and \cite{Wang:2012}.}
\end{figure}

\subsection{Comparison with the other {\it ab initio} results}

\begin{table}[b!]
  \caption{\label{tab_energy}Calculated ground-state energy of ${}^4$He with the decomposition of each cluster term.
The results in the section named "$V_{\rm low \ k}$" are calculated with the $V_{\rm low \ k}$ interaction
    derived from AV18 interaction with the sharp cutoff $\Lambda = 1.9$ fm$^{-1}$.
Also, the results in section "$V_{\rm SRG}$" are obtained with the SRG transformed chiral N$^{3}$LO $NN$ interaction
    at $\lambda_{\rm SRG} = 2.0$ fm$^{-1}$.
  All the energies are in MeV. See text for details.}
 \begin{ruledtabular}
  \begin{tabular}{llrrrr}
    &    Method                &      $E^{{\rm 1BC}} $   &     $E^{{\rm 2BC}}$   &     $E^{{\rm 3BC}}$   &   $E_{\rm g.s.}$ \\ \hline
$V_{\rm low \ k}$ &  UMOA    &    $62.60$  &  $-89.84$  & $-0.86$   & $-28.10$       \\
             & CCSD \cite{Hagen:2007}    &      &                               &                             &  $-28.9$      \\
             & FY \cite{Nogga:2004,Hagen:2007}      &    &        &   &  $-29.19(5)$ \\
$V_{\rm SRG}$ & UMOA    &    $53.50$  & $-80.47$ & $-0.76$ & $-27.73$         \\
             & IT-NCSM \cite{Roth:2011}    &     &                               &                             &     $-28.25(1)$   \\
    Exp.\cite{Wang:2012}       & & & & & $-28.30$ \\
  \end{tabular}
 \end{ruledtabular}
\end{table}


\begin{table}[b!]
  \caption{\label{tab_vlowk_r}Calculated point-nucleon radius of ${}^4$He with the decomposition of each cluster term.
See text for details. }
 \begin{ruledtabular}
  \begin{tabular}{lrrr}
    &   $\sqrt{\widetilde{r}^{2({\rm 1BC})}}$ & $\sqrt{\widetilde{r}^{2({\rm 2BC})}}$ &  $\sqrt{\widetilde{r}^{2}}$  \\ \hline
    $V_{\rm low \ k}$ &  $1.30$ & $0.20$   & $1.32$ \\
    $V_{\rm SRG}$ &  $1.40$ & $0.13$ & $1.41$  \\
    Exp.\cite{Alkhazov:1997}   & & & $1.49(3)$ \\
  \end{tabular}
 \end{ruledtabular}
\end{table}

After discussing the introduction of the one-body correlation operator in the UMOA works well,
in this section, we compare the present results with those from the other {\it ab inito} calculations.
For this purpose, we employ the $V_{\rm low \ k}$ potential derived from AV18 two-nucleon
 interaction at $\Lambda = 1.9$ fm$^{-1}$ throughout the comparison.
In addition, we also employ the SRG transformed chiral N$^{3}$LO $NN$ interaction~\cite{Entem:2003} with
 the momentum cutoff $\lambda_{\rm SRG} = 2.0$ fm$^{-1}$, which is widely used in recent
 {\it ab initio} calculations.

The results of the $^4$He ground-state energy are summarized in Table~\ref{tab_energy}
with the comparison to the other {\it ab inito} results.
In Table~\ref{tab_energy}, the calculated energies are shown with the decomposition of the one-body cluster (kinetic) term, $E^{{\rm 1BC}}$,
two-body cluster (interaction) term, $E^{{\rm 2BC}}$, the three-body cluster term, $E^{{\rm 3BC}}$, and the total energy,
$E_{\rm g.s.}$ for $^{4}$He, respectively (see Eq. (\ref{egs_3})).
All the results are calculated at $e_{\text{max}}=14$ and $\hbar\omega=20$ MeV.
For comparison, the results by the CCSD, FY, importance-truncated no-core shell model (IT-NCSM), and experiment
are taken from Refs. \cite{Hagen:2007}, \cite{Nogga:2004, Hagen:2007}, \cite{Roth:2011}, and  \cite{Wang:2012}, respectively.
We notice that the convergence with respect to the model-space size is confirmed.
The difference between $e_{{\rm max}} = 12$ and $14$ results is less than the order
 of 10 keV both for the $V_{\rm low \ k}$ and SRG transformed interactions.
As shown in Table~\ref{tab_energy}, the contribution of the three-body cluster term, $E^{\rm 3BC}$,
is much smaller than that of one- and two-body cluster terms, $E^{\rm 1BC}$ and $E^{\rm 2BC}$, respectively.
Therefore, our result for the energy practically converges with respect to the cluster expansion.
Moreover, the present result for the total energy, $E_{\text{g.s.}}$, is reasonably close to the other {\it ab initio} calculation energies.
The difference between them is comparable to the size of $E^{\rm 3BC}$.
In other words, the contributions of the truncated cluster terms can be approximated by the size of $E^{\rm 3BC}$ and
  the uncertainty of our energies can be roughly estimated from $E^{\rm 3BC}$.
In Fig.~\ref{4He_final}, we also summarize the comparison of various calculated energies.
The error bands of the UMOA energies are estimated from the size of the three-body cluster term corrections.

Our results for point-nucleon radius of $^{4}$He are listed in Table~\ref{tab_vlowk_r}.
It summarizes the calculated point-nucleon radii with the "one-body cluster
term " $\sqrt{\widetilde{r}^{2({\rm 1BC})}}$, "two-body cluster term" $\sqrt{\widetilde{r}^{2({\rm 2BC})}}$,
and the "one- and two-body cluster term" $\sqrt{\widetilde{r}^{2}}$, respectively (see Eq. (\ref{EffOp2})).
Caution that $\sqrt{\widetilde{r}^{2}} \neq \sqrt{\widetilde{r}^{2({\rm 1BC})}} + \sqrt{\widetilde{r}^{2({\rm 2BC})}}$,
and $\widetilde{r}^{2} = \widetilde{r}^{2({\rm 1BC})} + \widetilde{r}^{2({\rm 2BC})}$.
The calculation set up is same as in Table~\ref{tab_energy}.
The experimental value is take from Ref. \cite{Alkhazov:1997}.
 According to Table~\ref{tab_vlowk_r}, the contribution from the one-body cluster
 term, $\sqrt{\widetilde{r}^{2({\rm 1BC})}}$, is dominant.
Since the effects of the higher-body cluster terms can be expected to be smaller
 than that of one-body cluster term,
 the cluster expansion works well for the radius operator same as the case of the ground-state energy.
Moreover, the present result for point-proton radius, 1.41 fm, with the SRG softened N$3$LO chiral EFT $NN$ interaction~\cite{Entem:2003}
 with $\lambda_{\rm SRG}= 2.0$ fm$^{-1}$ is consistent with the in-medium SRG results with the same interaction \cite{Hergert:2016}.
Note that the calculated radius is much smaller than the experimental radius 1.49 fm.
 This observation is also true for the other {\it ab initio} calculations with $NN$ interactions only,
and is consistent with the fact of the obtained larger binding energy compared to the experimental binding energy.
From these demonstrations, we can conclude that the UMOA with the aid of the one-body correlation operator
gives fully-microscopic results as well as the other {\em ab initio} approaches at least in the ground-state properties of the $^4$He nucleus.

\section{\label{sec:sum}Summary}

We extend the former formalism of the UMOA accommodating only two-body correlations
by including additionally the one-body correlation operator.
As a demonstration, we have carried out numerical calculations
of the ground-state energy and point-nucleon radius for the $^{4}$He nucleus.
We can successfully reduce the $\hbar\omega$-dependence on calculated observables which exists in the earlier UMOA.
The present results for converged ground-state energy and radius are reasonably close to the results
from the other {\em ab initio} calculations with the same $NN$ interactions.
We find that the hierarchy of the cluster expansion is preserved
and the $\hbar\omega$-dependence can be removed order by order of the cluster expansion.
The decoupling of the $1p1h$ excitations from the $0p0h$ reference state
 is related to the HF mean-field to absorb the $\hbar\omega$-dependence.
The origin of the $\hbar\omega$-dependence in the former UMOA has also been discussed qualitatively.
Concerning the behavior of the total energy as a function of $\hbar\omega$,
the one-body kinetic term totally entails the positive slope,
while the two-body potential term causes the negative slope,
which results in the parabolic shape of the total energy.
The above discussion about the one-body kinetic energy
can also explain the negative slope of the radius as a function of $\hbar\omega$.

In this study, we only employ two-nucleon interactions without genuine and/or induced three-nucleon interactions.
For quantitative comparison with experimental data, the introduction of three-body forces is requisite.
For this purpose, the systematic improvement of the UMOA is necessary.
The next step to be done for this direction is to explicitly deal with the three-body cluster term
and to introduce the three-body correlation operator.
These implementations in the UMOA described above will open the way towards fully-microscopic description of nuclear structure
consistent with the other {\em ab-inito} methods.
As for the physical application, in this paper, we only focus on the ground-state property of the lightest doubly-closed nucleus, $^4$He,
as a test case of the implementation of the one-body correlation operator in the UMOA.
Application to heavier nuclei such as $^{ 16}$O and $^{40}$Ca is also interested.
The extension to the excited states can also be done, for example, with the method introduced in Ref.~\cite{Fujii:2004}.
Furthermore, the UMOA can be applied on the nuclear structure of open-shell nuclei
and gives information about the one-nucleon separation energy \cite{Fujii:2004, Fujii:2009}.
These studies are now on-going, and the results will be reported elsewhere in future publications.

\begin{acknowledgments}
The authors thank M.~Kohno, K.~Suzuki, H.~Kumagai, S.~Fujii, N.~Shimizu,
 B.~R.~Barrett, P.~Navr\'atil, and S.~R.~Stroberg
for many useful discussions.
This work was supported in part by JSPS KAKENHI Grant Number JP16J05707
 and by the Program for Leading Graduate Schools, MEXT, Japan.
This work was also supported in part by MEXT SPIRE and JICFuS (project IDs hp160211 and hp170230), and the CNS-RIKEN joint project for large-scale nuclear structure calculations.
\end{acknowledgments}

\appendix

\section{Decoupling equations \label{sec:dec_eq}}
In this appendix, we briefly explain how to determine the correlation operators used in the UMOA \cite{Suzuki:1982}.
Owing to the definition of the unitary operator $U$ as shown in Eq. (\ref{uni_op}),
 $\exp S^{(n)}$ appears only in the $n$- and higher-body cluster terms and does
 not affect cluster terms lower than $n$-body clusters
 (for instance, two- and higher many-body correlation operators do not show up in the one-body cluster terms of Eqs. (\ref{1b_h}) and (\ref{1b_int}).).
Therefore, we can sequentially solve correlation operators in the order from  the one- to $A$-body cluster terms.
In the following, we focus on the determination of the $n$-body correlation operator $S^{(n)}$, provided
 that the correlation operators with the rank lower than $n$,  $S^{(1)}$, $S^{(2)}$, $\cdots$, $S^{(n-1)}$, are already obtained.
The correlation operator $S^{(n)}$ is determined
 so that $\widetilde{H}^{(n)}$ has no matrix elements between the $0p0h$ and $npnh$ states.
For this purpose, we define the operators $P^{(n)}$ and $Q^{(n)}$
 projecting onto the space of $n$ particles occupying the orbits below
 and above the Fermi level, respectively.
Then, the decoupling condition between the target space $P^{(n)}$ and its complement $Q^{(n)}$ can be described as
\begin{align}
\label{dec_con}
 Q^{(n)}\widetilde{H}^{(n)}P^{(n)}
= P^{(n)}\widetilde{H}^{(n)}Q^{(n)}=0 .
\end{align}
Eq. (\ref{dec_con}) can be rewritten as
\begin{align}
& Q^{(n)}e^{-S^{(n)}}\widetilde{H}'^{(n)}e^{S^{(n)}}P^{(n)} \notag \\
 \label{dec_con2}
&\hspace{4em} =P^{(n)}e^{-S^{(n)}}\widetilde{H}'^{(n)}e^{S^{(n)}}Q^{(n)}=0
\end{align}
with
\begin{align}
  \widetilde{H}'^{(n)} &=
 e^{-(\sum_{i_{1}< \cdots < i_{n-1}}^n s_{i_{1}\cdots i_{n-1}})}\cdots
 e^{-(\sum_{i}^n s_{i})} \notag \\
&\hspace{-3em} \times
 \left(\sum_{i}^{n}h_{i}+\sum_{k=2}^{n}\sum_{i_{1}<\cdots
 <i_{k}}^{n}v_{i_{1}\cdots i_{k}}+
\sum_{k=2}^{n}\sum_{i_{1}<\cdots< i_{k}}^{n}w_{i_{1}\cdots
 i_{k}}\right) \notag \\
\label{prim_H}
& \times  e^{\sum_{i}^n s_{i}} \cdots
 e^{\sum_{i_{1}< \cdots < i_{n-1}}^n s_{i_{1}\cdots i_{n-1}}}.
\end{align}
For deriving Eq. (\ref{dec_con2}) from Eq. (\ref{dec_con}),
one can use the fact that
\begin{align}
  &Q^{(n)} \left(\sum_{i_{n}<\cdots < i_{n-1}}^{A}
\widetilde{w}_{i_{1}\cdots i_{n-1}} \right)P^{(n)} = 0, \\
 &Q^{(n)}\left(\sum_{i}^{n}\widetilde{h}_{i} +
 \sum_{k=2}^{n-1}\sum_{i_{1}<\cdots <i_{k}}^{n}
 \widetilde{v}_{i_{1}\cdots i_{k}}\right) P^{(n)} = 0,
\end{align}
 in Eqs. (\ref{nb_h}) and (\ref{nb_int}), since $P^{(n)}$ and $Q^{(n)}$ have no common states.
Thus,  $\widetilde{H}^{(n)}$ in Eq. (\ref{dec_con}) can be factorized  into $\widetilde{H}'^{(n)}$ and $e^{S^{(n)}}$.
$\widetilde{H}'^{(n)}$ can be calculated in a straightforward way, and
the problem can be cast into the determination of $S^{(n)}$.
According to Refs. \cite{Shavitt:1980,Westhaus:1981,Suzuki:1982},
it is known that one of the solutions can be expressed as
\begin{equation}
\label{Sn}
 S^{(n)} = \text{arctanh}(\omega^{(n)}-\omega^{(n)\dag})
\end{equation}
with the wave operator for n-particle states,
\begin{equation}
 \label{wn}
 \omega^{(n)}=\sum_{k=1}^{d}Q^{(n)}|\psi^{(n)}_{k}\>\<\widetilde{\phi}^{(n)}_{k}|P^{(n)}.
\end{equation}
Here, $d$ is the dimension of the $P^{(n)}$ space and
the bi-orthogonal state $\<\widetilde{\phi}^{(n)}_{k}|$ of $|\phi^{(n)}_{k}\>=P^{(n)}|\psi^{(n)}_{k}\>$
is defined to satisfy the following bi-orthonormal condition
\begin{equation}
 \label{bior}
 \<\widetilde{\phi}^{(n)}_{k}|\phi^{(n)}_{l}\>=\delta_{kl}.
\end{equation}
In Eq. (\ref{wn}), $|\psi^{(n)}_{k} \>$ is an eigenvector of
 the $n$-body Schr\"odinger equation
 in the $P^{(n)}+Q^{(n)}$ space,
\begin{equation}
 \label{shc_n}
 (P^{(n)}+Q^{(n)})\widetilde{H}'^{(n)}(P^{(n)}+Q^{(n)})|\psi^{(n)}_{k}\>
 = E_{k}|\psi^{(n)}_{k}\>,
\end{equation}
where $E_{k}$ is the $k$-th eigenvalue.
Moreover, $\exp S^{(n)}$ can also be expressed in terms of $\omega^{(n)}$ \cite{Suzuki:1994},
\begin{equation}
 \label{exp_omega}
  e^{S^{(n)}} =
  (1+\omega^{(n)}-\omega^{(n)\dag})(1+\omega^{(n)\dag}\omega^{(n)} +
  \omega^{(n)}\omega^{(n)\dag})^{-1/2}.
\end{equation}
Note that the solution of $\omega^{(n)}$ depends on the choice of a set of
$d$ eigenstates. In the present work, we choose $d$ eigenstates having the
largest overlap with the reference state.
This choice is reasonable as long as we consider only the ground state.

\section{One-body density matrix\label{Sec:density}}
Here, we summarize how to compute the one-body density matrix in the UMOA.
As found in usual textbooks (see, for example, Ref.~\cite{Ring:2004}),
the one-body density-matrix element
 $\gamma_{ba}$ is defined as
\begin{equation}
\label{density_1}
\gamma_{ba} \equiv  \<\Psi|n^{ab}|\Psi\>  = \<\Psi|c^{\dag}_{a}c_{b}|\Psi\>
\end{equation}
with the one-body operator $n^{ab}$ concerning to the single-particle states labeled by $a$ and $b$.
The one-body operator $n^{ab}$ can be transformed as
\begin{align}
  \gamma_{ba} &= \<\Phi|U^{\dag} n^{ab}U|\Phi\> \notag \\
  & = \<\Phi|\widetilde{n}^{ab}|\Phi\>
\end{align}
with the unitary-transformed one-body operator $\widetilde{n}^{ab} = U^{\dag}n^{ab}U$ and the reference state $|\Phi\>$.
Same as the Hamiltonian and radius operator discussed in the text, we apply the cluster expansion to this transformed one-body operator,
\begin{equation}
  \widetilde{n}^{ab} = \widetilde{n}^{ab(1)} + \widetilde{n}^{ab(2)} + \cdots.
\end{equation}
Here, one- and two-body cluster terms are defined as
\begin{equation}
  \widetilde{n}^{ab(1)} = \sum_{i} \widetilde{n}^{ab}_{i}, \quad
  \widetilde{n}^{ab(2)} = \sum_{i<j} \widetilde{n}^{ab}_{ij}
\end{equation}
with
\begin{align}
  \widetilde{n}^{ab}_{1} &= e^{-s_{1}} n^{ab}_{1} e^{s_{1}}, \\
  \widetilde{n}^{ab}_{12} &= e^{-s_{12}}(\widetilde{n}^{ab}_{1} + \widetilde{n}^{ab}_{2}) e^{s_{12}}
   - (\widetilde{n}^{ab}_{1} + \widetilde{n}^{ab}_{2}).
\end{align}
Then, the matrix element $\gamma_{ba}$ can also be expanded as
\begin{equation}
\gamma_{ba} = \sum_{\lambda \le \rho_{F}} \<\lambda|\widetilde{n}^{ab}_{1}|\lambda\> +
\frac{1}{2} \sum_{\lambda \mu \le \rho_{F}} \<\lambda\mu|\widetilde{n}^{ab}_{12} |\lambda\mu\>
 + \cdots.
\end{equation}
In the actual UMOA calculations, we keep the terms up to the two-body clusters.
With the aid of the one-body density matrix, the expectation value for one-body operators,
 $O = \sum_{ab}\<a|o|b\> c^{\dag}_{a}c_{b}$, can also be obtained by
 \begin{align}
   \<\Psi|O|\Psi\> &=   \sum_{ab}\<a|o|b\>\<\Psi|c^{\dag}_{a}c_{b}|\Psi\> = \sum_{ab}o_{ab}\gamma_{ba} \notag \\
   \label{tr-exp}
   & = {\rm Tr}(o\gamma)
 \end{align}
with the one-body matrix element $o_{ab} = \<a|o|b\>$.

\section{Expectation value of a contact interaction with respect to
the $(0s1/2)^{4}$ configuration \label{sec:toymodel}}
In this appendix, we derive the expectation value of the contact interaction $V_{\delta}$ introduced in Eq.~(\ref{eq:gaus}).
It is given by the normal ordered zero-body term with respect to the $(0s1/2)^{4}$ configuration:
\begin{equation}
  \<V_{\delta}\> = \sum_{JT} (2J+1) (2T+1) \<JT|V_{\delta}|JT\>
\end{equation}
with the $JT$-coupled two-body matrix element $\<JT|V_{\delta}|JT\>$.
Note that $|JT\>$ means the $(0s1/2)^{2}$ two-nucleon state with the total angular momentum $J$ and
 total isospin $T$.
Also $J + T$ has to be odd integer, because of the antisymmetrization of two-nucleon state.
To obtain the expectation value $\<V_{\delta}\>$,
all the terms we need are $\<10|V_{\delta}|10\>$ and $\<01|V_{\delta}|01\>$.
Applying the Talmi-Moshinsky transformation formula, they are expressed as
\begin{align}
  \<10|V_{\delta}|10\> &= \<\alpha|V_{\delta}|\alpha \>, \\
  \<01|V_{\delta}|01\> &= \<\beta| V_{\delta}|\beta \>
\end{align}
with the two-nucleon states $|\alpha\>$ and $|\beta\>$ in the relative coordinate.
Here, $|\alpha\>$($|\beta\>$) has the quantum numbers $(n, l_{\rm rel}, S, J_{\rm rel}, T) = (0, 0, 0, 0, 1)$
($(n, l_{\rm rel}, S, J_{\rm rel}, T) = (0, 0, 1, 1, 0)$),
where $n, l_{\rm rel}, S, J_{\rm rel}$, and $T$ are the nodal quantum number,
 relative orbital angular momentum, total spin, relative total angular momentum, and
 total isospin for the two-nucleon system, respectively.
 Since the HO radial wave function with $n = 0$ and $l_{\rm rel} = 0$ is just the Gaussian,
 these expectation values are written as
 \begin{align}
   \label{eq:rel1s0}
   \<\alpha|V_{\delta}|\alpha\> &= C_{^{1}S_{0}}\frac{8\sqrt{2}}{\sqrt{\pi}}
   \left( \frac{\hbar}{m\omega} \right)^{3/2} \Lambda_{\delta}^{6}
   I^{2}(\omega, \Lambda_{\delta}), \\
   \label{eq:rel3s1}
   \<\beta|V_{\delta}|\beta\> &= C_{^{3}S_{1}}\frac{8\sqrt{2}}{\sqrt{\pi}}
   \left( \frac{\hbar}{m\omega} \right)^{3/2} \Lambda_{\delta}^{6}
   I^{2}(\omega, \Lambda_{\delta})
 \end{align}
 with the integration
 \begin{equation}
   I(\omega, \Lambda_{\delta}) =  \int^{\infty}_{0} dx x^{2} \exp \left[
     -  \left(\frac{\hbar \Lambda_{\delta}^{2}}{m\omega } + 1 \right)
   x^{2}\right].
 \end{equation}
This integration can be done analytically and reads
\begin{equation}
  I(\omega, \Lambda_{\delta}) = \frac{\sqrt{\pi}}{4} \left(
    \frac{m\omega}{m\omega + \hbar \Lambda_{\delta}^{2}}
  \right)^{3/2}.
\end{equation}
Substituting this result into Eqs.~(\ref{eq:rel1s0}) and (\ref{eq:rel3s1}),
one can get
\begin{align}
  \label{eq:rel1s0_2}
  \<\alpha|V_{\delta}|\alpha\> &= C_{^{1}S_{0}}\sqrt{\frac{\pi}{2}}
  \left(
    \frac{\sqrt{\hbar m\omega} \Lambda_{\delta}^{2}}
    {m\omega + \hbar\Lambda_{\delta}^{2}}
  \right)^{3}, \\
  \label{eq:rel1s0_2}
  \<\beta|V_{\delta}|\beta\> &= C_{^{3}S_{1}}\sqrt{\frac{\pi}{2}}
  \left(
    \frac{\sqrt{\hbar m\omega} \Lambda_{\delta}^{2}}
    {m\omega + \hbar\Lambda_{\delta}^{2}}
  \right)^{3}.
\end{align}
Therefore, the expectation value can be obtained as
\begin{equation}
   \<V_{\delta}\>
   =  3 \sqrt{\frac{\pi}{2}} (C_{^{1}S_{0}} + C_{^{3}S_{1}})
  \left(
    \frac{\sqrt{\hbar m\omega} \Lambda_{\delta}^{2}}
    {m\omega + \hbar\Lambda_{\delta}^{2}}
  \right)^{3}.
\end{equation}
Since the $S$-wave scattering phase shift analysis implies that the $NN$ interaction
is attractive at low energies,
the low-energy constants $C_{^{1}S_{0}}$ and $C_{^{3}S_{1}}$ are usually negative values.
Then, the sign of $\<V_{\delta}\>$ is also negative.
To investigate the $\omega$-dependence of $\<V_{\delta}\>$, let us take the
derivative of $\<V_{\delta}\>$ with respect to $\omega$.
After the straightforward calculation, it is
\begin{align}
  & \frac{d\<V_{\delta}\>}{d\omega}
  = \frac{3}{2} \<V_{\delta}\>
  \frac{m\Lambda_{\delta}^{2}}
  {m\omega + \hbar\Lambda_{\delta}^{2}}
  \notag \\
  \label{eq:deri}
  & \times
  \left(\sqrt{\frac{\hbar}{m\omega}} + \frac{1}{\Lambda_{\delta}}
  \right)
  \left(\sqrt{\frac{\hbar}{m\omega}} - \frac{1}{\Lambda_{\delta}}
  \right).
\end{align}
From Eq.~(\ref{eq:deri}),  it is found that
the sign of the derivative is negative (positive) for $\hbar\omega < \hbar^{2}\Lambda_{\delta}^{2} / m$
($\hbar\omega > \hbar^{2}\Lambda_{\delta}^{2} / m$).
Thus, $\<V_{\delta}\>$ has a minimum at $\hbar\omega = \hbar^{2}\Lambda^{2}_{\delta}/m$.
Since $\hbar^{2}\Lambda_{\delta}^{2}/m$ specifies the energy scale of the
$NN$ interaction and is roughly the order of, at least,
the pion mass, this minimum is far from the $\hbar\omega$ values of 20-40 MeV in the
 present work.
Therefore, the $\hbar\omega$-dependence can be regarded as
monotonically decreasing in such a $\hbar\omega$ range.

\end{document}